\documentclass[preprint,aps,showpacs,amsmath,amssymb,prd,eqsecnum]{revtex4-1}
\usepackage{latexsym}
\usepackage{amsmath}
\usepackage{amsfonts}
\usepackage{amssymb}
\usepackage{array}

\allowdisplaybreaks

\newtheorem{thm}{Theorem}[section]

\newtheorem{lem}[thm]{Lemma}

\begin{document}
\title{Invariants of the heat equation for non-minimal operators}

\author{I. G. Moss and D. J. Toms}

\affiliation{ School of Mathematics and Statistics, Newcastle University, 
Newcastle Upon Tyne, NE1 7RU, UK}

\email{{ian.moss@newcastle.ac.uk},{david.toms@newcastle.ac.uk}}

\begin{abstract}
A special class of non-minimal operators which are relevant for quantum field theory is
introduced. The general form of the heat kernel coefficients of these operators on manifolds without 
boundary is described. New results are 
presented for the traces of the first two heat kernel coefficients for vector, Yang-Mills and perturbative
gravity. It is argued that non-minimal operators can be used to define gauge-fixing independent actions and
solve the conformal mode problem in quantum gravity.
\end{abstract}
\maketitle

\section{Introduction}{\label{genintro}}

The coefficients that arise in the asymptotic expansion of the heat kernel provide a very powerful tool in the analysis of divergences that arise in quantum field theory. This was first emphasized most explicitly by DeWitt~\cite{DeWittdynamical} who developed a now standard method for calculating the heat kernel coefficients for operators that are of interest to quantum field theory. (Independent work by mathematicians was given by Minakshisundaram and Pleijel~\cite{minakshisundaram1949some}. Later work by mathematicians can be found in \cite{gilkey1995invariance} and citations therein.) Subsequently the method was adapted and used to regularize effective Lagrangians~\cite{dowker1976effective} and path integrals~\cite{hawking1977zeta}. The method is now widely used and is covered in many reviews. (See~\cite{birrell1984quantum,barvinsky1985generalized,fulling1989aspects,avramidi2000heat,vassilevich2003heat,kirsten2010spectral,ParkerTomsbook} for example.) It is worth emphasizing that regularization methods that rely on the heat kernel are not restricted to field theories at one loop order, but can be applied to interacting fields at more than one loop as well. (See \cite{BunchParker,TomsPRDscalar} for two early references.) In addition the heat kernel expansion is of interest to a variety of mathematical problems. (See for example  \cite{gilkey1995invariance} and references therein.)

With the notable exception of Barvinsky and Vilkovisky~\cite{barvinsky1985generalized}, most of the references listed above deal with operators that are usually termed as minimal. A second order differential operator is called minimal if the second derivative part of the operator is comprised  solely of the Laplacian. Minimal operators are the easiest to deal with, but they form only a small class of second order operators. Non-minimal operators arise in gauge theories or gravity where gauge-fixing terms lead to additional second order derivative terms beyond the Laplacian. There are situations where the non-minimal character of the operator appears to be an essential feature.

The first situation is linearised Euclidean quantum gravity. The kinetic part of the Lagrangian density for linear perturbations $h_{\mu\nu}$ about a background Riemannian metric $g_{\mu\nu}$ with the DeWitt gauge-fixing function is \cite{DeWitt67,barvinsky1985generalized}
\begin{equation}
{\cal L}_K=\frac12 {\overline h}^{\mu\nu;\rho}h_{\mu\nu;\rho}
+\left(\frac{1}{\xi}-1\right)\,g^{\mu\nu}\overline h_{\mu\rho}{}^{;\rho}\overline h_{\nu\sigma}{}^{;\sigma},
\end{equation}
where $\xi$ is a gauge parameter and $\bar h_{\mu\nu}=h_{\mu\nu}-\frac12 g_{\mu\nu}h^\rho{}_\rho$.
The second order graviton operator obtained from this Lagrangian density is minimal if $\xi=1$. However, the
path integral for linearised gravity is badly behaved at $\xi=1$ due to a problem with the conformal mode 
$h_{\mu\nu}=\phi \,g_{\mu\nu}$. If we insert the conformal mode into the Lagrangian density we have
\begin{equation}
{\cal L}_K=\left(\frac{m-2}{2}\right)^2\left\lbrack\frac{1}{\xi}-\frac{2\,(m-1)}{ (m-2)}\right\rbrack(\nabla\phi)^2,
\end{equation}
where $m$ is the spacetime dimension. The action is positive definite only when $\displaystyle{\xi<\frac{(m-2)}{2(m-1)}}$, and never for the minimal case. In the minimal case we have to resort to some kind of contour rotation \cite{Gibbons1978141} or extra gauge-fixing \cite{PhysRevD.36.2342}
to define the path integral, although neither of these
fixes is particularly satisfactory.

Another reason for considering non-minimal operators arises from the construction of the effective action due to Vilkovisky~\cite{Vilkovisky1} and DeWitt~\cite{DeWitt2}.
In order to obtain an effective action which is independent of the gauge fixing, extra terms have to be added to the
field operators. The minimal operator may give misleading results and gauge condition dependent results. The Vilkovisky-DeWitt effective action
coincides with the standard effective action in the $\xi\to 0$ limit of the gauge parameter for theories such as Yang-Mills theory. The case of gravity is more complicated but it turns out that it is significantly easier to compute the Vilkovisky-DeWitt effective action in the case $\xi\rightarrow0$. (See \cite{FradkinTseytlin} or the pedagogical treatment in \cite{ParkerTomsbook} for full details.) We believe it is significant that
the Vilkovisky-DeWitt effective action for quantum gravity corresponds to a limit which has no conformal mode problem.

The third reason for considering non-minimal operators comes from boundary value problems in quantum gravity.
The gauged-fixed action has a residual BRST symmetry which we would like to apply also in the boundary
value problem for the quantum operators \cite{Moss:1989wu}. The BRST-invariant boundary value problem for minimal 
graviton operator does not have a well-defined heat kernel \cite{Avramidi:1997sh}, but there have been some 
indications that the heat kernel can be  defined for the boundary value problem with non-minimal 
operators \cite{Avramidi:1997hy}. This issue may be related to the conformal mode problem.

We will only consider manifolds without boundary. In the cases of interest to quantum field theory it 
appears that the non-minimal operator belongs to a special class of non-minimal operators. 
These are self-adjoint operators which
take the following form
\begin{equation}
\Delta^{i}{}_{j}=-\delta^i{}_jg^{\mu\nu}\nabla_\mu\nabla_\nu-
\zeta\,(P^{\mu\nu})^i{}_j\nabla_\mu\nabla_\nu+Q^i{}_j,
\end{equation}
where $\zeta$ is a real parameter and $\nabla :C^{\infty}(V)\to C^{\infty}(T M\otimes V)$ is a connection
for fields in a vector bundle $V$. The tensor $P^{\mu\nu}$ has two important properties, namely 
that $\nabla_\rho P^{\mu\nu}=0$ and
that the symbol of the non-minimal term $\hat P(x,p)= (P^{\mu\nu})^i{}_j\hat p_\mu\hat p_\nu$, 
with unit vectors $\hat p_\mu$, is a projection, i.e.
\begin{equation}
\hat P^2=\hat P.
\end{equation}
The parameter $\zeta$ controls the size of the non-minimal part of the
operator, and for $\zeta>-1$, the operator is of elliptic form and the heat kernel
has a well-defined asymptotic expansion,
\begin{equation}
K(x,x,\tau)\sim (4\pi \tau)^{-m/2}\sum_{n=0}^\infty E_n(x,\Delta) \tau^{n}.
\end{equation}
We will show that the heat kernel coefficients for the special class of operators
belong to the algebra of invariants constructed from the curvature, $Q$ and $P^{\mu\nu}$. 
Furthermore, $E_n$ contains at most $3n$ occurrences of the
tensor $P^{\mu\nu}$. We will also give explicit formulae for the traces of $E_1$ and $E_2$ for vector
and tensor fields.

An early field theory calculation of the $E_1$ coefficient for a non-minimal vector field was given in \cite{parker1984renormalization} (but with a typographical error that was corrected in \cite{HKLT}.) Another early paper on non-minimal electromagnetic fields is \cite{endo1984gauge}. One of the methods that we use in the present paper is an extended version of what was used in \cite{parker1984renormalization}. A comprehensive and systematic method for calculating the integrated trace of the heat kernel coefficients for non-minimal operators was undertaken by Barvinsky and Vilkovisky~\cite{barvinsky1985generalized}. These authors describe an elegant extension of the method of DeWitt~\cite{DeWittdynamical} and apply it to many of the operators of interest to quantum field theory. However by working out only the integrated trace of the relevant coefficients, all terms which are total derivatives are lost. In addition, the spacetime dimension is restricted to four and, unlike the case for minimal operators, the heat kernel coefficients for non-minimal operators have an explicit dependence on the spacetime dimension even before the trace is taken. An important paper by Branson, Gilkey and Fulling~\cite{gilkey1991heat} established that for the special case of an operator of the form $a\delta d+ bd\delta$ acting on forms the traced integrated heat kernel coefficients can be related to those of the minimal operator. For the more general operator $a\delta d+ bd\delta-E$ acting on forms the first two traced and integrated coefficients ${\rm Tr}\,E_0$ and ${\rm Tr}\,E_1$ were found. (See also \cite{fulling1992kernel,fulling1992kernel2,branson1994heat}.)

A more general case of the unintegrated and untraced heat kernel coefficients was undertaken by Gusynin and Kornyak~\cite{gusynin1997computation,gusynin1999complete}. They considered an operator of the general form
\begin{equation}\label{GK}
-\delta^{\mu}{}_{\nu}\nabla^2+a\nabla^{\mu}\nabla_{\nu}+Q^{\mu}{}_{\nu},
\end{equation}
where $a$ is some constant and $Q^{\mu}{}_{\nu}$ involves no derivatives. Here $\nabla_\mu$ is some covariant derivative that may involve a gauge as well as a Christoffel connection. This operator is of relevance to Yang-Mills theory. General expressions for the first three heat kernel coefficients were obtained on a manifold of general dimension. Our results are in broad agreement with those of \cite{gusynin1997computation,gusynin1999complete} in this special case \eqref{GK} as we will discuss below.

The aim of the present paper is to present a more general set of results for the non-minimal operator than those of \cite{gusynin1997computation,gusynin1999complete}, and to describe how a method based on the local momentum space method of Bunch and Parker~\cite{BunchParker} may be used. This method is of interest in its own right in quantum field theory because it provides an expansion of the Feynman Green function that can be used in quantum field theory calculations beyond one-loop order. (See for example the review in \cite{ParkerTomsbook}.) However we do not discuss such applications in the present paper. In addition we analyze the heat kernel coefficients for the operator of interest to quantum gravity which does not take the form \eqref{GK}. We will discuss detailed checks of our results against previously known special cases as outlined above. The method that we describe is straightforward although the calculations are extremely lengthy. The most tedious parts of the calculations were done using {\tt Cadabra}~\cite{Peeters1,Peeters2}.

\section{Basic formalism}{\label{djtintro}}

Consider a generic Bose field $\varphi^i(x)$ on a spacetime manifold $M$. Here $i$ represents any type of indices, for example, vector or tensor, or gauge group. Suppose that we use $\Delta^{i}{}_{j}$ to be the relevant self-adjoint differential operator for the field $\varphi^i$. We choose a Riemannian spacetime metric, take the spacetime dimension to be $m$, and adopt the curvature conventions of Misner, Thorne and Wheeler \cite{MTW}. The heat kernel $K^{i}{}_{j}(x,x';\tau)$ is a solution to 
\begin{equation}
\Delta^{i}{}_{j}K^{j}{}_{k}(x,x';\tau)=-\frac{\partial}{\partial\tau}K^{i}{}_{k}(x,x';\tau),\label{djt1.1}
\end{equation}
with the boundary condition
\begin{equation}
K^{i}{}_{j}(x,x';\tau=0)=\delta^{i}{}_{j}\delta(x,x').\label{djt1.2}
\end{equation}
Here $\delta(x,x')$ is the biscalar Dirac delta distribution. For the rest of this section, and whenever there is no confusion, we will omit the internal indices
$i,j$.

The importance of the heat kernel is that under fairly general assumptions it admits an asymptotic expansion as $\tau\rightarrow 0$ of the form
\begin{equation}
K(x,x;\tau)\sim(4\pi\tau)^{-m/2}\sum_{k=0}^{\infty}\tau^kE_k(x).\label{djt1.3}
\end{equation}
The coefficients $E_k(x)$ are the heat kernel coefficients that are local expressions determined solely by the form of the operator $\Delta$. (Note that we do not consider any contributions from a possible boundary here.)

The method that we will use here makes use of the Green function for the operator $\Delta$ rather than the heat kernel directly. There is a simple relationship between the two. Normally the Green function is defined as the solution to 
\begin{equation}\label{djt1.4}
\Delta G(x,x')=I\,\delta(x,x'),
\end{equation}
where $I$ is the unit matrix $\delta^i{}_j$. This Green function is the analytic continuation of the normal Feynman Green function (or propagator) to imaginary time.  It proves convenient to define an auxiliary Green function $G(x,x';s)$ as the solution to
\begin{equation}\label{djt1.5}
(\Delta-sI)G(x,x';s)=I\,\delta(x,x').
\end{equation}
The usual Green function $G(x,x')$ in \eqref{djt1.4} clearly is related to the auxiliary Green function $G(x,x';s)$ by
\begin{equation}\label{djt1.6}
G(x,x')=G(x,x';s=0).
\end{equation}
The relation between the auxiliary Green function and the heat kernel is
\begin{equation}\label{djt1.7}
G(x,x';s)=\int\limits_{0}^{\infty}d\tau\;e^{s\tau}\,K(x,x';\tau),
\end{equation}
which can be recognized as a one-sided Laplace transform \cite{DuffNaylor}. The inverse of this, giving the heat kernel in terms of the auxiliary Green function, can be obtained as 
\begin{equation}\label{djt1.8}
K(x,x';\tau)=\int\limits_{c-i\infty}^{c+i\infty}\frac{ds}{2\pi i}\;e^{-s\tau}\;G(x,x';s).
\end{equation}
Here $c$ is chosen to be a real constant smaller than the lowest eigenvalue of the differential operator $\Delta$ and the contour is closed in the right hand side of the complex $s$-plane. It is easily verified, using (\ref{djt1.4}), that the heat kernel obeys (\ref{djt1.1}). The boundary condition \eqref{djt1.2} follows by using the expansion of the Green function in terms of eigenfunctions of the operator $\Delta$.

We will be interested in the case where 
\begin{equation}
\Delta=A^{\mu\nu}\,\partial_\mu\partial_\nu +B^{\mu}\,\partial_\mu +C\label{djt1.9}
\end{equation}
for some matrix coefficients $A^{\mu\nu},\ B^{\mu}$ and $C$. Without loss of generality we can assume $A^{\nu\mu}=A^{\mu\nu}$. The method that we will adopt makes use of the local momentum space approach of Bunch and Parker~\cite{BunchParker}, which is a special case of the symbol calculus of Seeley (see e.g. \cite{gilkey1995invariance}), to calculate the auxiliary Green function in \eqref{djt1.5}. The first step is to expand the matrix coefficients
about a selected point $x'$, and for this we introduce the convenient multi-index notation $\alpha=(\alpha_1,\dots ,\alpha_m)$, and 
\begin{equation}
|\alpha|=\alpha_1+\dots+\alpha_m,\quad
\alpha!=\alpha_1!\dots\alpha_m!,\quad
y^\alpha=(y^1)^{\alpha_1}\dots (y^m)^{\alpha_m}.
\end{equation}
Introduce normal coordinates at the point $x$ and let $y^\mu=x^{\prime\mu}-x^{\mu}$.  The coefficients in \eqref{djt1.9} can all be expanded about $y^\mu=0$,
\begin{eqnarray}
A^{\mu\nu}(x')&=&A_0^{\mu\nu}(x)+\sum_{|\alpha|\ge 2}A^{\mu\nu}{}_\alpha(x) y^\alpha,\label{djt1.10}\\
B^\mu(x')&=&\sum_{|\alpha|\ge 1}B^{\mu}{}_\alpha(x) y^\alpha,\label{djt1.11}\\
C(x')&=&C_0(x)+\sum_{|\alpha|\ge 1}C_\alpha(x) y^\alpha\;.\label{djt1.12}
\end{eqnarray}
These are not the most general possibilities for these expansions, but are sufficient to deal with the cases that arise in the present paper. The absence of a linear term in $y^\mu$ in \eqref{djt1.10} is a consequence of the fact that in the examples we will deal with $A^{\mu\nu}$ depends only on the spacetime metric whose expansion in Riemann normal coordinates has the first non-trivial term quadratic in $y^\mu$. Similarly, the absence of a zeroth order term in \eqref{djt1.11} arises because $B^\mu$ involves the connection whose Riemann normal coordinate expansion begins at order $y^\mu$. For the case where these conditions are not met, and the more general results that ensue, see \cite{toms2011quadratic}.

The next step is to Fourier transform the equation for the Green function (\ref{djt1.7}). The Fourier transform of a bitensor $F(x,x')$
in the appropriate class defines the symbol $\sigma F(x,p)$,
\begin{equation}
\sigma F(x,p)=\int d^m x' e^{-i(x-x')^\mu p_\mu} F(x,x').\label{djt1.13}
\end{equation}
The inverse transform is then
\begin{equation}
F(x,x')=\int \frac{d^m p}{ (2\pi)^m} e^{i(x-x')^\mu p_\mu} \sigma F(x,p).\label{djt1.14}
\end{equation}
For example, if we take the original operator $\Delta$, then the symbol is obtained by replacing $\partial_\mu$
with $ip_\mu$,
\begin{equation}
\sigma\Delta=-A^{\mu\nu}p_\mu p_\nu+iB^\mu p_\mu+C.
\end{equation}
For polynomials in the momentum $p_\mu$ like this, we will use $\sigma\Delta_n$ to denote the term of order $n$.
Because of the Fourier transform, this method is called the local momentum space expansion in quantum field theory~\cite{BunchParker}. In functional
analysis, the bitensors belong to the class of pseudo-differential operators.

The aim now is to make use of the expansions \eqref{djt1.10}--\eqref{djt1.12} in \eqref{djt1.5}. There is a useful relation for the symbol of a product of two operators $F(x,x')$ and $H(x.x')$, which can be obtained by inserting the inverse transforms \eqref{djt1.14},
\begin{equation}
\sigma(FH)=\sum_\alpha \frac{1}{\alpha !}D^\alpha(\sigma F)\partial_\alpha(\sigma H),\label{sigmaFH}
\end{equation}
where $D$ is the momentum derivative,
\begin{equation}
D^\alpha=(-i)^{|\alpha|}\left(\frac{\partial}{ \partial p_1}\right)^{\alpha_1}\dots\left(\frac{\partial}{ \partial p_m}\right)^{\alpha_m}.
\end{equation}
We apply this to the product $G(\Delta-sI)$, noting that the left and right inverses of the operator $\Delta-sI$ are equivalent
because we have assumed that the operator $\Delta$ is self-adjoint. The Fourier transform then gives
\begin{equation}
\sigma\left( G(\Delta-sI)\right)= I.
\end{equation}
This can be solved recursively by setting $\sigma G=G_0+G_1+\dots$, where $G_n$ can be thought of as having
order $p^{-2-n}$ for large $p$. After using \eqref{sigmaFH},
\begin{equation}
\sum_n\sum_{S(\alpha,j,l)=n}\frac{1}{\alpha !}D^\alpha G_j\partial_\alpha\sigma(\Delta-sI)_l\sim I
\end{equation}
where $S(\alpha,j,l)=|\alpha|+j-l+2$. The symbol $\sim$ has been used to remind us that the series obtained from \eqref{sigmaFH} will match the right hand side of this equation term by term, but might not be convergent. The first term in the series gives
\begin{equation}
G_0(A^{\mu\nu}p_\mu p_\nu+sI)=-I.\label{defgo}
\end{equation}
Separating of the $\alpha=0$ terms from the rest of the series gives the general formula for $G_n$,
\begin{equation}
G_n=-\sum_{S(\alpha,j,l)=n}\frac{1}{\alpha !}D^\alpha G_j\,\partial_\alpha\sigma\Delta_l\,G_0.\label{gn}
\end{equation}
The $sI$ term has disapeared because it is always eliminated by the $\partial_\alpha$. 

The $n=1$ term vanishes from the assumptions used in defining the normal coordinate expansions \eqref{djt1.10}--\eqref{djt1.12}.
After expanding the multi-index notation into single-index notation, the next term $G_2$ is
\begin{equation}
G_2=-\frac{\partial^2 G_0}{\partial p_\alpha\partial p_\beta}A^{\mu\nu}{}_{\alpha\beta}p_\mu p_\nu G_0
-\frac{\partial G_0}{\partial p_\alpha}B^\mu{}_\alpha p_\mu G_0-G_0 C G_0.\label{g2}
\end{equation}
Proceding further gives $G_4$ in terms of $G_0$ and $G_2$,
\begin{eqnarray}
G_4&=&\frac{\partial^4 G_0}{\partial p_\alpha\partial p_\beta\partial p_\gamma\partial p_\delta}
A^{\mu\nu}{}_{\alpha\beta\gamma\delta}p_\mu p_\nu G_0
-\frac{\partial^2 G_2}{\partial p_\alpha\partial p_\beta}A^{\mu\nu}{}_{\alpha\beta}p_\mu p_\nu G_0\nonumber\\
&&-\frac{\partial^3 G_0}{\partial p_\alpha\partial p_\beta\partial p_\gamma}
B^\mu{}_{\alpha\beta\gamma} p_\mu G_0-\frac{\partial G_2}{\partial p_\alpha}B^\mu{}_\alpha p_\mu G_0
+\frac{\partial^2 G_0}{\partial p_\alpha\partial p_\beta}C_{\alpha\beta}G_0.\label{g4}
\end{eqnarray}
The evaluation of $G_n$ for larger $n$ and a general operator soon becomes impractical due to the large number of terms. As we will show the terms indicated here are sufficient to evaluate the first three terms in the heat kernel expansion. The term $G_3$ which we do not display cannot contribute to the heat kernel coefficients.

The heat kernel is related to the Green function by \eqref{djt1.8} and can therefore be expanded as a series
$K_0+K_1+\dots$. If we take $x^{\prime\mu}\rightarrow x^{\mu}$ and then substitute the inverse transform \eqref{djt1.13} we 
obtain a formula for $K_n(x,x,\tau)$,
\begin{equation}
K_n(x,x;\tau)=\int\frac{d^mp}{(2\pi)^m} \int\limits_{c-i\infty}^{c+i\infty}\frac{ds}{2\pi i}\;e^{-s\tau}\, G_n(x,p;s).\label{djt1.31}
\end{equation}
Note that $G_n$ as defined in \eqref{defgo} and \eqref{gn} obeys the scaling relation
\begin{equation}
G_n(x,\tau^{-1/2}p;s/\tau)=\tau^{1+n/2} G_n(x,p;s).\label{djt1.33}
\end{equation}
We use the scaling relation to scale $\tau$ out of the integral in \eqref{djt1.31} and obtain
\begin{equation}
K_n=\tau^{(n-m)/2}\int\frac{d^mp}{(2\pi)^m}\int\limits_{c-i\infty}^{c+i\infty}\frac{ds}{2\pi i}\,e^{-s}\,G_n(x,p;s).\label{djt1.32a}
\end{equation}
From this we read off the heat kernel coefficient $E_k$ defined in \eqref{djt1.3},
\begin{equation}
E_k=(4\pi)^{m/2}\int\frac{d^mp}{(2\pi)^m}\int\limits_{c-i\infty}^{c+i\infty}\frac{ds}{2\pi i}\,e^{-s}\,G_{2k}(x,p;s).\label{djt1.32}
\end{equation}
We are able to obtain the untraced heat kernel coefficients from this formula, but most of the results we will derive later are for the traces
of these coefficients. (The untraced heat kernel coefficients are useful in some calculations that will be given elsewhere. The untraced expressions for $G_n$ are needed to obtain even the traced heat kernel coefficients as the results in \eqref{g2} and \eqref{g4} show.)

So far our considerations have been reasonably general and have followed a familiar path. We stated in the introduction that,
in quantum field theory applications, $A^{\mu\nu}$ takes a special form. We will assume that,
at the origin of normal coordinates,
\begin{equation}
A^{\mu\nu}=-I\delta^{\mu\nu}-\zeta P^{\mu\nu},\label{djt1.34}
\end{equation}
for some constant $\zeta$ and matrix $P^{\mu\nu}$. In the special case where $\zeta=0$ the operator that describes the theory is minimal and the heat kernel coefficients are well-known 
\cite{gilkey1975spectral,barvinsky1985generalized,fulling1989aspects,gilkey1995invariance,avramidi2000heat,vassilevich2003heat,kirsten2010spectral,ParkerTomsbook}. Suppose that we define a normalised symbol of $P^{\mu\nu}$ by
\begin{equation}
\hat P=P^{\mu\nu}\hat p_\mu \hat p_\nu.\label{djt1.35}
\end{equation}
(So $\hat p^\mu \hat p_\mu=1$.) We consider operators where $\nabla_\rho P^{\mu\nu}=0$ and $\hat P$ is a projection operator
\begin{equation}
\hat P^2=\hat P.\label{djt1.36}
\end{equation}
This latter condition allows the flat-spacetime momentum-space Feynman Green function to be found because it allows the operator in \eqref{defgo} to be inverted in closed form. We will justify that this is the case for spin one and spin two fields in the next section. It is then easy to show that the solution for $G_0$ in \eqref{defgo} is
\begin{equation}
G_0(x,p;s)=(p^2-s)^{-1}I-\zeta(p^2-s)^{-1}\lbrack(1+\zeta)p^2-s\rbrack^{-1}\,p_\mu p_\nu\, P^{\mu\nu}.\label{djt1.37}
\end{equation}
By combining this with the general formula for $G_n$ \eqref{gn} and the heat-kernel coefficient \eqref{djt1.32}, we see that
the general term in $E_n$ resembles $PXPY\dots P$, where $X,Y,\dots$ are derivatives of the matrix coefficients $A^{\mu\nu}$, $B^\mu$
and $C$. This can be made more quantitative by introducing an order, which is essentially the inverse mass (or equivalently the length) dimension in the quantum
field theory applications. We set
\begin{equation}
{\rm ord}\ A^{\mu\nu}{}_\alpha=|\alpha|,\quad 
{\rm ord}\ B^\mu{}_\alpha=1+|\alpha|,\quad
{\rm ord}\ C_\alpha=2+|\alpha|.
\end{equation}
It follows from \eqref{gn} that ${\rm ord} \,G_n=n$ and then by the integration \eqref{djt1.32} that
\begin{lem}\label{lemma}
The  heat kernel coefficient $E_n$ is a polynomial in $A^{\mu\nu}{}_\alpha$, $B^\mu{}_\alpha$, $C_\alpha$ with
${\rm ord}\,E_n=2n$ and each term has at most $2n$ factors of $P^{\mu\nu}$.
\end{lem}

We have reduced the problem of the evaluation of the heat kernel coefficients to one that is purely algorithmic.
Although the method is very straightforward, the implementation involves considerable algebraic complexity so that the results are best performed by computer. The {\tt Cadabra} program~\cite{Peeters1,Peeters2} has been used for most of the results, and some have been
checked using {\tt Maple}. The lengthy technical details will be omitted here. (A more pedagogical description will be presented elsewhere~\cite{DJTheatkernelbook}.) It can be shown that under the assumptions that we have described in the previous section resulting in the expression for $G_0$ in \eqref{djt1.37} the general result for ${\rm Tr}E_0$ is
\begin{equation}
{\rm Tr}E_0={\rm Tr}I-\frac{1}{m}\left\lbrack 1-(1+\zeta)^{-m/2}\right\rbrack {\rm Tr}P,\label{djt2.1}
\end{equation}
where $P=P^\mu{}_\mu$. (The untraced expression is simply \eqref{djt2.1} with the traces omitted.)

In the higher order coefficients, the momentum integrals reduce to a scalar factor multiplied by
a symmetric tensor $t_{\mu_1\dots\mu_n}$ defined by
\begin{equation}
t_{\mu_1\dots\mu_n}=(4\pi)^{m/2}\int\frac{d^m p}{ (2\pi)^m} \hat p_{\mu_1}\dots \hat p_{\mu_n} e^{-p^2}.
\end{equation}
The integral has been normalised so that the tensor has unit trace. Explicitly, we have
$t_{\mu_1\mu_2}=\delta_{\mu_1\mu_2}/m$ and
\begin{eqnarray}
t_{\mu_1\dots\mu_4}&=&\frac{3}{ m(m+2)} \delta_{(\mu_1\mu_2}\delta_{\mu_3\mu_4)},\\
t_{\mu_1\dots\mu_6}&=&\frac{15}{ m(m+2)(m+4)} 
\delta_{(\mu_1\mu_2}\delta_{\mu_3\mu_4}\delta_{\mu_5\mu_6)}.
\end{eqnarray}
The general result for $E_1$ becomes
\begin{eqnarray}
{\rm Tr}E_1&=&b_1\,t_{\mu\nu\rho\sigma}{\rm Tr}A^{\mu\nu\rho\sigma}
+b_2\,t_{\mu\nu\rho\sigma\alpha\beta}{\rm Tr}P^{\alpha\beta}A^{\mu\nu\rho\sigma}\nonumber\\
&&+b_3\,t_{\mu\nu\rho\sigma}{\rm Tr}P^{\alpha\mu}A^{\nu\rho\sigma}{}_\alpha
+b_4\,t_{\mu\nu\rho\sigma}{\rm Tr}P^{\mu\alpha}P^{\nu\beta}A^{\rho\sigma}{}_{\alpha\beta}
\nonumber\\
&&+b_5\,{\rm Tr}B^{\prime\mu}{}_\mu+b_6\,t_{\mu\nu\rho\sigma}{\rm Tr}P^{\mu\nu}B^{\prime\rho\sigma}
+b_7\,{\rm Tr}P^{\mu\nu}B^\prime_{\mu\nu}\nonumber\\
&&+b_8\,t_{\mu\nu\rho\sigma}{\rm Tr}P^{\mu\nu}P^{\rho\alpha}B^{\prime\sigma}{}_{\alpha}
+b_9\,{\rm Tr}C+b_{10}\,{\rm Tr}P C,\label{djt2.2}
\end{eqnarray}
where
\begin{equation}
B^{\prime\mu}{}_\alpha=B^{\mu}{}_\alpha-2A^{\mu\beta}{}_{\alpha\beta}.
\end{equation}
The coefficients that appear in front of the various terms have been evaluated but are given by lengthy expressions
which we will omit. However, some features of these terms are worthy of mention, so consider the example
\begin{eqnarray}
b_8&=&{\frac { 2\left( 4+2\,\zeta +m\,  \zeta  \right)}{m
-2}}\left\{ (1+\zeta)^{-m/2}-1\right\}+{\frac {4m\, \zeta }{m-2}}.
\end{eqnarray}
Note that the cases $m=2$, $\zeta=0$ and $\zeta\to\infty$ are all special. 
The limit $m\rightarrow2$ exists, and
\begin{equation}
b_8\to\frac{4}{ \zeta}\ln(1+\zeta)-2\frac{2+\zeta}{ 1+\zeta}.
\end{equation}
We have checked by a careful evaluation of the coefficients that the case $m=2$ agrees with the procedure of 
taking the $m\rightarrow2$ limit of our general results. The limit $\zeta\to0$ corresponds to the minimal operator, 
and we find that $b_8\to0$ in this limit as we would expect. The limit $\zeta\to\infty$, as used in the Vilkovisky-DeWitt
effective action, gives a divergent result for $b_8$. The validity of the Vilkovisky-DeWitt effective action 
requires cancellations between some of the terms in the heat kernel coefficients, and in addition there are usually extra terms in the effective action from ghost fields that are not considered here. The examples investigated below will all
have these cancellations and give finite results for the heat kernel coefficients as $\zeta\to\infty$. As mentioned earlier we will not look at the direct physical applications in this paper.

\section{Covariant expansions}\label{covariant}

Recall that the operator of interest has a covariant form,
\begin{equation}
\Delta=-g^{\mu\nu}\nabla_\mu\nabla_\nu-
\zeta P^{\mu\nu}\nabla_\mu\nabla_\nu+Q\label{op},
\end{equation}
where $\nabla_\rho P^{\mu\nu}=0$ and the gauge indices have been suppressed. 
The covariant derivative includes both the Levi-Civita connection for $g_{\mu\nu}$
and the gauge connection $W_\mu$, 
with any tangent-space indices on the fields included alongside internal gauge indices.
Since the commutator of two covariant derivatives produces a curvature term, we
need only consider the case where $P^{\mu\nu}$ is symmetric in the spacetime indices.

The first step in applying the general formalism is to write the operator
(\ref{op}) in the form \eqref{djt1.9} and identify the expressions for $A^{\mu\nu}$,
$B^\mu$ and $C$. Some simplification occurs if we make a preliminary
rescaling by the determinant $|g(x)|$ to a new operator $\Delta'$ of the form,
\begin{equation}
\Delta'=|g(x)|^{1/4}\Delta|g(x')|^{-1/4}.
\end{equation}
This has no effect on the heat kernel coefficients. (This has been checked in the examples described below by explicit calculation.) In coordinates, the
new operator becomes
\begin{equation}
\Delta'=-\left(\partial_\mu-\frac12\omega_{,\mu}+W_\mu\right)
\left(g^{\mu\nu}+\zeta P^{\mu\nu}\right)
\left(\partial_\nu+\frac12\omega_{,\nu}+W_\nu\right)
+Q
\end{equation}
where $\omega=\ln|g(x)|^{-1/2}$. We can read off the matrix coefficients of
the operator $\Delta'$ at an
arbitrary point $x'$,
\begin{eqnarray}
A^{\mu\nu}&=&-g^{\mu\nu}-\zeta P^{\mu\nu}\\
B^\mu&=&\partial_\nu A^{\mu\nu}+A^{\mu\nu}W_\nu+W_\nu A^{\mu\nu}\\
C&=&Q+(\partial_\nu A^{\mu\nu})\left(W_\nu+\frac12\omega_{,\nu}\right)
+\frac14A^{\mu\nu}(2\omega_{,\mu\nu}-\omega_{,\mu}\omega_{,\nu})\nonumber\\
&&+A^{\mu\nu}\partial_\mu W_\nu+W_\mu A^{\mu\nu}W_\nu.
\end{eqnarray}
The next step is to find the coefficients in the normal coordinate expansions of
$A^{\mu\nu}$, $B^\mu$ and $C$ as defined in \eqref{djt1.10}--\eqref{djt1.12}. We require the normal coordinate
expansions of the metric, the gauge field, the endomorphism Q and the covariantly 
constant tensor $P^{\mu\nu}$.
The normal coordinate expansion for the metric is a standard result (see \cite{Moss:1999wq} for example),
\begin{eqnarray}
g_{\mu\nu}(x')&=&\delta_{\mu\nu}-\frac{1}{3}y^\alpha y^\beta\,R_{\mu\alpha\nu\beta} -\frac{1}{6} y^\alpha y^\beta y^\gamma\,R_{\mu\alpha\nu\beta;\gamma}\nonumber\\
&&+y^\alpha y^\beta y^\gamma y^\delta\Big(-\frac{1}{20}R_{\mu\alpha\nu\beta;\gamma\delta} +\frac{2}{45}R_{\mu\alpha\beta\lambda}R_{\nu\gamma\delta}{}^{\lambda}\Big)+\cdots,\label{3.1.5}\\
g^{\mu\nu}(x')&=&\delta^{\mu\nu}+\frac{1}{3}\,y^\alpha y^\beta\,R^{\mu}{}_{\alpha}{}^{\nu}{}_{\beta} +\frac{1}{6}\, y^\alpha y^\beta y^\gamma\, R^{\mu}{}_{\alpha}{}^{\nu}{}_{\beta;\gamma}\nonumber\\
&&+y^\alpha y^\beta y^\gamma y^\delta\Big(\frac{1}{20}\,R^{\mu}{}_{\alpha}{}^{\nu}{}_{\beta;\gamma\delta} +\frac{1}{15}\,R^{\mu}{}_{\alpha\beta\lambda}R^{\nu}{}_{\gamma\delta}{}^{\lambda} \Big)+\cdots,\label{3.1.6}\\
\omega(x')&=&1+\frac16 R_{\alpha\beta}y^\alpha y^\beta+
\frac{1}{12}R_{\alpha\beta;\gamma}y^\alpha y^\beta y^\gamma\nonumber\\
&&+\left(\frac{1}{40}R_{\alpha\beta;\gamma\delta}
+\frac{1}{180}R^{\mu}{}_{\alpha\beta\nu}R^{\nu}{}_{\gamma\delta\mu}\right)
y^\alpha y^\beta y^\gamma y^\delta+\dots.
\end{eqnarray}
When expanding the gauge connection, we choose a basis for the internal space and
parallel transport it from $x$ to $x'$. The details can be found in \cite{Moss:1999wq}, and the result is that
\begin{eqnarray}
W_\mu(x')&=&-\frac12F_{\mu\alpha}y^\alpha-\frac13 F_{\mu\alpha;\beta}\,y^\alpha y^\beta\nonumber\\
&&-\left(\frac18F_{\mu\alpha;\beta\gamma}+\frac{1}{18}R_{\mu\alpha\nu\beta}F_\gamma{}^\nu\right)
y^\alpha y^\beta y^\gamma+\dots.\label{wnc}
\end{eqnarray}
The semi-colon denotes a covariant derivative with both the Levi-Civita and gauge connection.
The normal coordinate expansions of the matrices $Q$ and $P^{\mu\nu}$ can be obtained by expanding 
the covariant derivatives order by order in $y^\mu$,
\begin{eqnarray}
Q(x')&=&Q+y^\alpha\, Q_{;\alpha}+\frac{1}{2}\,y^\alpha y^\beta\,Q_{;\alpha\beta}+\cdots,\label{3.1.8}\\
P^{\mu\nu}(x')&=&P^{\mu\nu}
+\frac13R^{(\mu}{}_{\alpha\sigma\beta}P^{\nu)\sigma}y^\alpha y^\beta
+\frac16R^{(\mu}{}_{\alpha\sigma\beta;\gamma}P^{\nu)\sigma}y^\alpha y^\beta y^\gamma
\nonumber\\
&&+\left(\frac{1}{20}R^{(\mu}{}_{\alpha\sigma\beta;\gamma\delta}P^{\nu)\sigma}
+\frac{1}{36}R^{\mu}{}_{\alpha\beta\rho}R^{\nu}{}_{\gamma\delta\sigma}P^{\rho\sigma}\right.\nonumber\\
&&\left.\qquad+\frac{7}{180}R^{\sigma}{}_{\alpha\beta\rho}R^{\mu}{}_{\gamma\delta\sigma}P^{\nu)\sigma}\right)
 y^\alpha y^\beta y^\gamma y^\delta+\dots.\label{3.1.88}
\end{eqnarray}
All terms on the right-hand side of these equations are evaluated at the origin of normal
coordinates. The expansion has been done up to the order we require for the calculation of $E_2$,
but these expansions could be continued to arbitrary order in principle. In these expansions we have converted all of the spacetime indices on $Q$ and $P^{\mu\nu}$ that correspond to the fields to tangent space indices with the appropriate vierbeins. In some of the expressions in the particular examples described later we do not do this as it provides a useful check on the detailed calculations.

With these expansions it is easy to verify that the assumptions that we
made concerning the expansions of $A^{\mu\nu}$ and $B^{\mu}$ are met. (Specifically,
there is no term in the expansion of $A^{\mu}$ that is linear in $y^\alpha$, and the expansion
of $B^\mu$ begins at order $y^\alpha$.) Furthermore, the matrix coefficients $A^{\mu\nu}{}_\alpha$ etc
have at most one factor of $P^{\mu\nu}$ and their order is related to the orders of 
$R_{\mu\nu\rho\sigma}$, $Q$ and $P^{\mu\nu}$ if we set
\begin{equation}
{\rm ord}\,R_{\mu\nu\rho\sigma}=
{\rm ord}\,F_{\mu\nu}={\rm ord}\,Q=2,\qquad {\rm ord}\,\nabla=1,
\qquad {\rm ord}\,P^{\mu\nu}=0.
\end{equation}
The earlier result Lemma~\ref{lemma} then implies
\begin{thm}\label{lem1}
The general coefficient $E_n(x,\Delta)$ for the special class of operators is a covariant polynomial in $R_{\mu\nu\rho\sigma}$, 
$Q$ and $P^{\mu\nu}$ and their derivatives with ${\rm ord}\,E_n=2n$ containing at most $3n$ factors of $P^{\mu\nu}$.
\end{thm}
We can always choose an orthonormal frame at the point $x$, and then we find
in the following section that $P^{\mu\nu}$ consists of a series of Kronecker delta tensors, 
some of them mixing spacetime and internal tangent bundle indices. 
After substituting for $P^{\mu\nu}$, the heat kernel coefficient will consist of terms which have contractions 
of $R_{\mu\nu\rho\sigma}$, $Q$ and their derivatives.

As an example, the explicit coefficients that we will need for $E_1$ are
\begin{eqnarray}
A^{\mu\nu}&=&-\delta^{\mu\nu}-\zeta P^{\mu\nu},\label{cova}\\
A^{\mu\nu}{}_{\alpha\beta}&=&-\frac13 R^\mu{}_\alpha{}^\nu{}_\beta
-\frac13\zeta R^{(\mu}{}_{\alpha\rho\beta}P^{\nu)\rho}.
\end{eqnarray}
The remaining coefficients are expressed in terms of $A^{\mu\nu}$ for simplicity,
\begin{eqnarray}
B^\mu{}_\alpha&=&2A^{\mu\nu}{}_{\alpha\nu}-\frac12\left\{A^{\mu\nu},F_{\nu\alpha}\right\},\\
C&=&Q+\frac16A^{\mu\nu}R_{\mu\nu},\label{covc}
\end{eqnarray}
where $\{F,G\}$ is used for the anti-commutator $FG+GF$. Higher order terms for specific cases  are given in the next section.

\begin{table}[htb]
\caption{\label{table3}The sub-coefficients in the traced heat kernel coefficient ${\rm Tr}\,E_1$ for 
the special class of non-minimal operators in spacetime dimension
$m$ and $u=(1+\zeta)^{-m/2}$. $\zeta$ is defined in \eqref{op}. }
\centering
\begingroup
\renewcommand{\arraystretch}{1.5}
\begin{tabular}{ll}
\hline\noalign{\smallskip}
Term&Expression\\
\noalign{\smallskip}\hline\noalign{\smallskip}
$a_1$&$\displaystyle\frac16$\\
$a_2$&$-1$\\
$a_3$&$\displaystyle-\frac{u-1}{ m}$\\
$a_4$&$\displaystyle-{\frac { \left( m+2 \right)  \left( m\, \zeta  -
2\,m+4\,\zeta +10 \right) (u-1)}{ 12\left( m-2 \right) m\, \left( m-1
 \right) }}+{\frac {  \zeta  \, \left( -9\,m+{m}^{2
}+2 \right) }{6 \left( m-2 \right) m\, \left( m-1 \right) }}
$\\
$a_5$&$\displaystyle{\frac { \left( 8+{m}^{2} \zeta +4\,\zeta 
 \right) (u-1)}{4 \left( m-2 \right)  \left( m-1 \right) {m}^{2}}}+{\frac {
\zeta }{ \left( m-2 \right) m\, \left( m-1 \right) }}
$
\\
$a_6$&$\displaystyle{\frac { 2\left( 4+2\,\zeta +m\,  \zeta  \right) (u-1)}{m
-2}}+{\frac {4m\, \zeta }{m-2}}
$
\\
\noalign{\smallskip}\hline
\end{tabular}
\endgroup
\end{table}

The covariant $E_1$ coefficient can be obtained by substituting the normal coordinate expansions 
\eqref{cova}-\eqref{covc} into the general expression \eqref{djt2.2} for the $E_1$ coefficient. The appendix
explains how some of the terms can be simplified by using the symmetry under tetrad rotations. It is possible to
replace most of the $P^{\mu\nu}$ terms by a trace over spacetime indices $P=P^\mu{}_\mu$. 
The traced $E_1$ coefficient is then
\begin{eqnarray}
{\rm Tr}\,E_1&=&a_1R\, {\rm Tr}(I)+a_2 {\rm Tr}\,Q+a_3 {\rm Tr}\,PQ
+a_4R\,{\rm Tr}\,P+a_5R\,{\rm Tr}\,P^2\nonumber\\
&&+a_6t_{\mu\nu\rho\sigma}{\rm Tr}\,P^{\mu[\nu}P^{\alpha]\rho}P^{\beta\sigma}F_{\alpha\beta},\label{tre1}
\end{eqnarray}
where the coefficients are given in table \ref{table3}. Note that these results are valid when $P^{\mu\nu}$
is symmetric in the spacetime indices.

\section{Some applications}\label{applications}

In this and the following sections we will apply the general methods from the previous section 
to a number of examples of direct physical interest. The calculations in this section on the $E_1$
coefficient have been done in two different ways as a check on the computer algebra programs 
which have been employed. The first method involves substitution into the general result \eqref{tre1},
which was obtained using Maple. The second method repeats the general steps, with some
modifications detailed below, using a {\tt Cadabra} program~\cite{Peeters1,Peeters2}. The results are summarised in table
\ref{table1}. Agreement between the two methods gives us some confidence in the results 
for the $E_2$ coefficients in the next section, where using a general result is impractical.

\begin{table}[htb]
\caption{\label{table1}The heat kernel coefficient ${\rm Tr}E_1=c_{1}\,R+c_{2} {\rm Tr}\,Q$ 
for vectors, $1-$forms and rank two symmetric tensors.
The spacetime dimension is $m$ and $u=(1+\zeta)^{-m/2}$ There are two ${\rm Tr}\,Q$ terms for
tensors which are given in section~\ref{gravity}.}
\centering
\begingroup
\renewcommand{\arraystretch}{1.5}
\begin{tabular}{lll}
\hline\noalign{\smallskip}
Example&$c_{1}$&$c_{2}$\\
\noalign{\smallskip}\hline\noalign{\smallskip}
 Vector&
$\displaystyle \,{\frac { \left( m+6+m\zeta \right) u}{6m}}+
{\frac { \left( m+2 \right)  \left( m-3 \right) }{6m}}$
&
$\displaystyle -{\frac {u}{m}}-{\frac {m-1}{m}}$
\\
$1-$form&
$\displaystyle \frac{(1+\zeta)u}{ 6}+\frac{m-7}{ 6}$
&
$\displaystyle -{\frac {u}{m}}-{\frac {m-1}{m}}$
\\
Tensor&
$\displaystyle{\frac { \left( {m}^{2}+{m}^{2}\zeta+6\,m\,\zeta+6\,m+24 \right) u}{6m}}
+{\frac {{m}^{3}-{m}^{2}-12\,m-48}{12m}}
$
&
see section \ref{gravity}
\\
\noalign{\smallskip}\hline
\end{tabular}
\endgroup
\end{table}

\subsection{Real vector field}\label{real vector}

The operator that describes the quantized spin-1 vector field takes the general form
\begin{equation}
\Delta^{\mu}{}_{\nu}=-\delta^{\mu}_{\nu}\,\nabla^2-\zeta \,\nabla^\mu\nabla_\nu+Q^{\mu}{}_{\nu}.\label{3.1.1}
\end{equation}
The general field indices ($i$ and $j$) of the general results are simply vector spacetime indices in this case that are raised and lowered with the spacetime metric $g_{\mu\nu}$ as usual. Here $\zeta$ is treated as a constant and arises in quantum field theory from a gauge fixing condition of the Landau-Feynman type. $Q^{\mu}{}_{\nu}$ is a general second rank tensor field that typically involves just the Ricci tensor in the case of the Maxwell field. If the vector field is massive then there will be an additional term $m^2\delta^{\mu}{}_{\nu}$ where $m$ is the mass. We will not specify the details of $Q^{\mu}{}_{\nu}$ here. Additionally we do not use the vierbein to relate the spacetime indices on $Q$ to those in the tangent space. As mentioned, this provides an independent check on our results.

The first step in applying the general results is to write the operator \eqref{3.1.1} in the form of \eqref{djt1.9} and identify the expressions for $A^{\mu\nu},B^\mu$ and $C$. This is done simply by expanding out the covariant derivatives in terms of the ordinary ones and comparing coefficient of the derivatives that occur. We do not employ the rescaling by $|g|^{1/4}$ which we used in the previous
section, so now the results are
\begin{eqnarray}
\big(A^{\mu\nu}\big)^{\lambda}{}_{\tau}&=&-g^{\mu\nu}\,\delta^{\lambda}_{\tau}-\frac{1}{2}\zeta\big( g^{\lambda\mu}\,\delta^{\nu}_{\tau}+g^{\lambda\nu}\,\delta^{\mu}_{\tau}\big),\label{3.1.2}\\
\big(B^{\mu}\big)^{\lambda}{}_{\tau}&=&-2\,g^{\mu\nu}\,\Gamma^{\lambda}_{\nu\tau}+g^{\alpha\beta}\,\Gamma^{\mu}_{\alpha\beta}\,\delta^{\lambda}_{\tau}-\zeta \,g^{\mu\lambda}\,\Gamma^{\nu}_{\nu\tau},\label{3.1.3}\\
\big(C\big)^{\lambda}{}_{\tau}&=&Q^{\lambda}{}_{\tau}+g^{\alpha\beta}\,\Gamma^{\gamma}_{\alpha\beta}\,\Gamma^{\lambda}_{\gamma\tau}-g^{\alpha\beta}\,\Gamma^{\lambda}_{\beta\gamma}\,\Gamma^{\gamma}_{\alpha\tau}-g^{\alpha\beta}\,\Gamma^{\lambda}_{\alpha\tau,\beta}\nonumber\\
&&\qquad-\zeta \,g^{\lambda\sigma}\,\Gamma^{\nu}_{\nu\tau,\sigma}.\label{3.1.4}
\end{eqnarray}

The next step is to identify the coefficients in the normal coordinate expansions of $A^{\mu\nu},B^\mu$ and $C$ as defined in \eqref{djt1.10}--\eqref{djt1.12}. In the real vector field case this just requires the normal coordinate expansion of the metric, inverse metric and Christoffel symbols. The relevant expressions for the metric are given in \eqref{3.1.5}, and for the Christoffel symbol
\begin{eqnarray}
\Gamma^{\lambda}_{\mu\nu}(x')&=&-\frac{1}{3}y^\alpha\big( R^{\lambda}{}_{\mu\nu\alpha} +R^{\lambda}{}_{\nu\mu\alpha}\big)-\frac{1}{12} y^\alpha y^\beta \big( 2R^{\lambda}{}_{\mu\nu\alpha;\beta} +2R^{\lambda}{}_{\nu\mu\alpha;\beta}  \nonumber\\
&&+R^{\lambda}{}_{\alpha\nu\beta;\mu} +R^{\lambda}{}_{\alpha\mu\beta;\nu} - R_{\mu\alpha\nu\beta}{}^{;\lambda}  \big)\nonumber\\
&&+y^\alpha y^\beta y^\gamma\Big( -\frac{1}{20} R^{\lambda}{}_{\mu\nu\gamma;\alpha\beta} -\frac{1}{20} R^{\lambda}{}_{\nu\mu\gamma;\alpha\beta} -\frac{1}{40} R^{\lambda}{}_{\beta\nu\gamma;\mu\alpha}\nonumber\\
&&-\frac{1}{40} R^{\lambda}{}_{\beta\mu\gamma;\nu\alpha}- \frac{1}{40} R^{\lambda}{}_{\beta\nu\gamma;\alpha\mu} -\frac{1}{40} R^{\lambda}{}_{\beta\mu\gamma;\alpha\nu} +\frac{1}{40} R_{\mu\beta\nu\gamma}{}^{;\lambda}{}_{\alpha}\nonumber\\
&& +\frac{1}{40} R_{\mu\beta\nu\gamma}{}^{;}{}_{\alpha}{}^{\lambda} +\frac{1}{45}R^{\lambda}{}_{\alpha\mu\sigma}R_{\nu\beta\gamma}{}^{\sigma} +\frac{1}{45}R^{\lambda}{}_{\alpha\nu\sigma}R_{\mu\beta\gamma}{}^{\sigma}\nonumber\\
&&  -\frac{4}{45}R^{\lambda}{}_{\alpha\beta\sigma}R_{\nu\gamma\mu}{}^{\sigma} -\frac{4}{45}R^{\lambda}{}_{\alpha\beta\sigma}R_{\mu\gamma\nu}{}^{\sigma} -\frac{1}{45}R^{\lambda}{}_{\sigma\mu\beta}R_{\nu\alpha\gamma}{}^{\sigma}\label{3.1.7}\\
&& -\frac{1}{45}R^{\lambda}{}_{\sigma\nu\beta}R_{\mu\alpha\gamma}{}^{\sigma} +\frac{2}{45}R^{\lambda}{}_{\mu\beta\sigma}R_{\nu\alpha\gamma}{}^{\sigma} +\frac{2}{45}R^{\lambda}{}_{\nu\beta\sigma}R_{\mu\alpha\gamma}{}^{\sigma}\Big)+\cdots.\nonumber
\end{eqnarray}
All curvature terms are evaluated at the origin of Riemann normal coordinates where the metric tensor reduces to $\delta_{\mu\nu}$. 
Unlike in the previous section, we use the normal coordinate frame for the internal vector indices  so that the expansion for 
$Q^{\lambda}{}_{\tau}$ turns out to be 
\begin{eqnarray}
Q^{\lambda}{}_{\tau}(x')&=&Q^{\lambda}{}_{\tau}+y^\alpha\, Q^{\lambda}{}_{\tau;\alpha}\nonumber\\
&&+\frac{1}{2}\,y^\alpha y^\beta\,\big( Q^{\lambda}{}_{\tau;\alpha\beta}+\frac{1}{3}\,R^{\lambda}{}_{\alpha\sigma\beta}Q^{\sigma}{}_{\tau} -\frac{1}{3}\,R^{\sigma}{}_{\alpha\tau\beta}Q^{\lambda}{}_{\sigma}\big)+\cdots,\label{3.1.8b}
\end{eqnarray}
with the usual understanding that all terms on the right hand side are evaluated at the origin of Riemann normal coordinates. We do not need all of the terms in the expansions that are shown here to calculate $E_0$ and $E_1$, but they will be needed in the next section when we calculate  $E_2$.

With these expansions again it is easy to verify that the assumptions that we made concerning the expansions of $A^{\mu\nu}$ and $B^\mu$ above are met. (Specifically, there is no term in the expansion of $A^{\mu\nu}$ that is linear in $y^\alpha$, and the expansion of $B^\mu$ begins at order $y^\alpha$.) The explicit coefficients that we will need for $E_1$ are
\begin{eqnarray}
\big(A_{0}^{\mu\nu}\big)^{\lambda}{}_{\tau}&=&-\delta^{\lambda}_{\tau}\delta^{\mu\nu}-\frac{1}{2}\,\zeta\big(\delta^{\mu\lambda}\delta^{\nu}_{\tau}+\delta^{\nu\lambda}\delta^{\mu}_{\tau}\big),\label{3.3.11sa}\\
\big(A^{\mu\nu}{}_{\alpha\beta}\big)^{\lambda}{}_{\tau}&=&-\frac{1}{3}\delta^{\lambda}_{\tau}R^{\mu}{}_{\alpha}{}^{\nu}{}_{\beta}-\frac{1}{6}\,\zeta\big( \delta^{\nu}_{\tau}R^{\lambda}{}_{\alpha}{}^{\mu}{}_{\beta} + \delta^{\mu}_{\tau}R^{\lambda}{}_{\alpha}{}^{\nu}{}_{\beta}\big),\label{3.3.11sb}\\
\big(B^{\mu}{}_{\alpha}\big)^{\lambda}{}_{\tau}&=&\frac{2}{3}\delta^{\lambda}_{\tau}R^{\mu}{}_{\alpha} -\frac{2}{3}R_{\alpha\tau}{}^{\lambda\mu}-\frac{2}{3}R_{\alpha}{}^{\mu\lambda}{}_{\tau} +\frac{1}{3}\,\zeta\,\delta^{\mu\lambda}\,R_{\tau\alpha},\label{3.3.11se}\\
\big( C_0\big)^{\lambda}{}_{\tau}&=&Q^{\lambda}{}_{\tau}+\frac{1}{3}R^{\lambda}{}_{\tau} +\frac{1}{3}\,\zeta\,R^{\lambda}{}_{\tau},\label{3.3.12sa}
\end{eqnarray}
The expansions needed for $E_2$ are given in an appendix.

We can identify $P^{\mu\nu}$ in (\ref{djt1.34}) from (\ref{3.3.11a}) as
\begin{equation}
\big(P^{\mu\nu}\big)^{\lambda}{}_{\tau}=\frac{1}{2}\,
\big(\delta^{\mu\lambda}\delta^{\nu}{}_{\tau}+\delta^{\nu\lambda}\delta^{\mu}{}_{\tau}\big).\label{3.3.13}
\end{equation}
We then have $\hat P$ in (\ref{djt1.35}) given by
\begin{equation}
\big(\hat P\big)^{\lambda}{}_{\tau}=\hat p^\lambda \hat p_\tau,\label{3.3.14}
\end{equation}
and it is easy to see that our assumption in \eqref{djt1.36} holds. All of the assumptions that we made in deriving the general results for $E_0$ and $E_1$ are satisfied in the case of a real vector field. It is now a straightforward, yet algebraically tedious matter, to use the results given here to evaluate the expressions given in \eqref{djt2.1} and \eqref{djt2.2}. The result for the trace of $E_0$ turns out to be the same as before
\begin{equation}
{\rm Tr}\,E_0=m-1+(1+\zeta)^{-m/2}.\label{3.3.16}
\end{equation}
(The untraced $E_0$ is a multiple of the identity.) For $E_1$ we find
\begin{equation}
{\rm Tr}\,E_1=c_{1}R+c_{2}{\rm Tr}\,Q.\label{3.3.17}
\end{equation}
where $c_1$ and $c_2$ are given in table \ref{table1}. (The untraced expression for $E_1$ is a special case of the result given in \ref{appuntraced}.)
In the limit where we take $\zeta=0$, so that the operator becomes minimal, we recover the standard result \cite{gilkey1975spectral,gilkey1995invariance}
\begin{equation}
{\rm Tr}E_1=\frac{1}{6}R\,{\rm Tr}\,I-{\rm Tr}\,Q.\label{3.3.21}
\end{equation}
This is true regardless of the spacetime dimension.
Finally, our results apply to $1-$forms with operator
\begin{equation}
\Delta=\delta d+(1+\zeta)\,d \delta+Q,
\end{equation}
where $d$ is the exterior derivative with adjoint $\delta$. We have to make the replacement 
$Q_\mu{}^\nu\to Q_\mu{}^\nu+R_\mu{}^\nu$ to recover the form of the vector operator
\eqref{3.1.1}, and this gives the $1-$form coefficients in table \ref{table1}. These agree with \cite{branson1994heat}. 

\subsection{Yang-Mills field}\label{yang mills}

We now consider the application of our results to non-Abelian, or Yang-Mills, gauge theories. We will use the conventions of \cite{toms1983yangmills}. The covariant form of the operator is (with gauge indices suppressed)
\begin{equation}
\Delta^{\mu}{}_{\nu}=-\delta^{\mu}_{\nu} D^2-\zeta \,D^\mu D_\nu+R^{\mu}{}_{\nu}-2\,F^{\mu}{}_{\nu}.\label{3.3.22}
\end{equation}
Here $D_\mu=\nabla_\mu+A_\mu$ is the gauge covariant derivative, the tangent space connection sits in $\nabla_\mu$ and 
the Yang-Mills gauge field $A_\mu$ has been separated out. The gauge coupling constant has been absorbed into the gauge field. In terms of the usual gauge fields $A^{a}_{\mu}$, where $a,b,c,\ldots$ represent the Lie algebra indices, we have $A_\mu=gA^{a}_{\mu}T_{a}$ and the Yang-Mills field strength $F_{\mu\nu}=gF^{a}_{\mu\nu}T_{a}$ with $T_a$ the anti-hermitian generators for the Lie algebra of the gauge group. We let $f_{abc}$ represent the structure constants and restrict attention to gauge groups whose Killing metric is the Kronecker delta (so that there is no distinction between upper and lower group indices) and that have totally antisymmetric structure constants $f_{abc}$. If we write out the operator in \eqref{3.3.22} with the gauge connection expanded out we have
\begin{eqnarray}
\Delta^{a\mu}{}_{b\nu}&=&-\big(\delta^{\mu}_{\nu}\,g^{\alpha\beta}+\zeta\,g^{\mu\alpha}\delta^{\beta}_{\nu}\big)
\big\lbrack\delta^{a}_{b}\nabla_\alpha\nabla_\beta-gf^{a}{}_{bc}A^{c}_{\beta}\nabla_\alpha
-gf^{a}{}_{bc}A^{c}_\alpha\nabla_\beta\nonumber\\
&&-gf^{a}{}_{bc}\nabla_\alpha A^{c}_{\beta} +g^2f^{a}{}_{cd}f^{c}{}_{be}A^{d}_{\alpha} A^{e}_{\beta}\big\rbrack +Q^{a\mu}{}_{b\nu}.\label{3.3.23}
\end{eqnarray}
For the Yang-Mills field we have
\begin{equation}
Q^{a\mu}{}_{b\nu}=\delta^{a}_{b}R^{\mu}{}_{\nu}+2gf^{a}{}_{bc}F^{c\mu}{}_{\nu},\label{3.3.24}
\end{equation}
but we will obtain results valid for arbitrary $Q^{a\mu}{}_{b\nu}$ here before specializing to Yang-Mills.

Our general results require the operator to be written in the general form of \eqref{djt1.9}. In the present case the general index $i$ stands for a pair of indices $(a,\mu)$. Expanding out the spacetime covariant derivatives will generate terms like those for the real vector field in Sec.~\ref{real vector}. In fact it is easy to see that
\begin{equation}
\big(A^{\alpha\beta}\big)^{a\mu}{}_{b\nu}=\delta^{a}{}_{b}\big(A^{\alpha\beta}_{{\rm vec}}\big)^{\mu}{}_{\nu},\label{3.3.25}
\end{equation}
where $\big(A^{\alpha\beta}_{{\rm vec}}\big)^{\mu}{}_{\nu}$ was the real vector field result in \eqref{3.1.2}. The operator $P^{\mu}{}_{\nu}$ has the special form
with $\hat P^2=\hat P$ as before. The relevant expansion coefficients defined in \eqref{djt1.10} are simply obtained from those in \eqref{3.3.11a}--\eqref{3.3.11d} by multiplying by $\delta^{a}{}_{b}$.

The results for $\big(B^\alpha\big)^{a\mu}{}_{b\nu}$ and $\big(C\big)^{a\mu}{}_{b\nu}$ can be read from \eqref{3.3.23}.
In order to obtain the normal coordinate expansions we use \eqref{3.1.5}, \eqref{3.1.6}, \eqref{3.1.7} and
 the expansions of the gauge field  \eqref{wnc}. The matrix coefficients we require for $E_1$ are found to be
\begin{eqnarray}
\big(B^{\alpha}{}_{\beta}\big)^{a\mu}{}_{b\nu}&=&\delta^{a}_{b}\, \big(B^{\alpha}{}_{\beta\,{\rm vec}}\big)^{\mu}{}_{\nu}-g\,\delta^{\mu}_{\nu}\,f^{a}{}_{bc}F^{c\alpha}{}_{\beta}\nonumber\\
&&\qquad-\frac{1}{2}\,\zeta\,g\,\delta^{\alpha}_{\nu}\,f^{a}{}_{bc}F^{c\mu}{}_{\beta}-\frac{1}{2}\,\zeta\,g\,\delta^{\alpha\mu}\,f^{a}{}_{bc}\,F^{c}_{\nu\beta}.\label{3.3.31}
\end{eqnarray}
and
\begin{equation}
\big(C_0\big)^{a\mu}{}_{b\nu}=\delta^{a}_{b}\, \big(\widetilde{C}_{0\,{\rm vec}}\big)^{\mu}{}_{\nu} +\frac{1}{2}\,\zeta\,g\,f^{a}{}_{bc}\,F^{c\mu}{}_{\nu} + Q^{a\mu}{}_{b\nu}.\label{3.3.32}
\end{equation}
Here $\big(B^\alpha_{{\rm vec}}\big)^{\mu}{}_{\nu}$ is the real vector field result and 
$\big(\widetilde{C}_{0\,{\rm vec}}\big)^{\mu}{}_{\nu}$ is the real vector field result given in \eqref{3.3.12a} 
with the omission of the term in $Q^{\mu}{}_{\nu}$.

As already explained, because of \eqref{3.3.25} we have $\big(P^{\alpha\beta}\big)^{a\mu}{}_{b\nu} =\delta^{a}_{b}\big(P^{\alpha\beta}_{{\rm vec}}\big)^{\mu}{}_{\nu} $, and hence
\begin{equation}
\big(G_0\big)^{a\mu}{}_{b\nu}=\delta^{a}_{b}\,\big(G_{0\,{\rm vec}}\big)^{\mu}{}_{\nu},\label{3.3.29}
\end{equation}
where $\big(G_{0\,{\rm vec}}\big)^{\mu}{}_{\nu}$ is the real vector field Green function given from \eqref{djt1.37} and \eqref{3.3.13}.
After some calculation, the result for the trace of $E_1$ turns out to be
\begin{equation}
{\rm Tr}\,E_1= c_{1} d_G\,R+c_{2}{\rm Tr}\,Q,\label{3.3.34}
\end{equation}
where the trace is over the group and tangent space indices and
\begin{eqnarray}
c_{1}&=&\frac{1}{6}(m+\zeta)+\frac{1}{6m}(m+m\zeta+6)\lbrack(1+\zeta)^{-m/2}-1\rbrack.\label{3.3.35a}\\
c_{2}&=&-1+\frac{1}{m}\lbrack1-(1+\zeta)^{-m/2}\rbrack.\label{3.3.35b}
\end{eqnarray}
(The untraced expression is given in \ref{appuntraced}.) In the special case of pure Yang-Mills theory where $Q^{a\mu}{}_{b\nu}$ is given by \eqref{3.3.24} we have
\begin{equation}
{\rm Tr}\,E_1=\frac16\left((1+\zeta)^{1-m/2}+(m-7)\right)\,d_G\,R.\label{3.3.36b}
\end{equation}
In the case of $m=4$, this agrees with standard results \cite{parker1984renormalization,HKLT}, and for $d_G=1$ we recover the
$1-$form result \cite{gilkey1991heat}.

\subsection{Spin-2 gravity}\label{gravity}

For gravity we will let $g_{\mu\nu}$ be the background spacetime metric and let $h_{\mu\nu}$ denote the quantum fluctuation about the background. If we choose $h_{\mu\nu}$ as our basic fields, then the natural metric on the space of fields is the DeWitt metric
\begin{equation}
{\mathcal G}^{\mu\nu\,\rho\sigma}=\frac{1}{2}(g^{\mu\rho}g^{\nu\sigma}+g^{\mu\sigma}g^{\nu\rho}-g^{\mu\nu}g^{\rho\sigma}).\label{3.4.1}
\end{equation}
The inverse of the DeWitt metric only exists in spacetimes whose dimension is other than 2. It is convenient to treat the lower indices
as the internal components corresponding to the index $i$ in the generic field $\varphi^i$. 
If this is done, the relevant operator for the spin-2 field is
\begin{equation}
\big(\Delta\big)_{\mu\nu}{}^{\lambda\tau}=- \big(\Sigma^{\alpha\beta}\big)_{\mu\nu}{}^{\lambda\tau}\nabla_\alpha\nabla_\beta +Q_{\mu\nu}{}^{\lambda\tau},\label{3.4.2}
\end{equation}
where
\begin{equation}
\big(\Sigma^{\alpha\beta}\big)_{\mu\nu}{}^{\lambda\tau}=
g^{\alpha\beta}\,\delta^{\lambda\tau}_{\mu\nu}+\zeta\big(\delta^{\alpha\lambda}_{\mu\nu}\,g^{\beta\tau} +\delta^{\alpha\tau}_{\mu\nu}\,g^{\beta\lambda}-\delta^{\alpha\beta}_{\mu\nu}\,g^{\lambda\tau}\big) .\label{3.4.3}
\end{equation}
We have introduced
\begin{equation}
\delta^{\alpha\beta}_{\mu\nu}=\frac{1}{2}\big(\delta^{\alpha}_{\mu}\,\delta^{\beta}_{\nu} + \delta^{\alpha}_{\nu}\,\delta^{\beta}_{\mu}\big),\label{3.4.4}
\end{equation}
which is the identity operator on the space of symmetric second rank tensors. It is important to remember that the DeWitt metric must be used to raise and lower field indices, rather than simply using the spacetime metric. We will keep $Q_{\mu\nu}{}^{\lambda\tau}$ general here, but record that in the case of Einstein gravity with a cosmological constant (see Eq.~(2.62) of \cite{HKLT} for example)
\begin{eqnarray}
Q_{\mu\nu}{}^{\lambda\tau}&=&R^{\lambda}{}_{\mu\nu}{}^{\tau}+R^{\lambda}{}_{\nu\mu}{}^{\tau}-\frac{1}{2}\big(\delta^{\lambda}_{\mu}R^{\tau}_{\nu}+\delta^{\tau}_{\mu}R^{\lambda}_{\nu} +\delta^{\lambda}_{\nu}R^{\tau}_{\mu}+\delta^{\tau}_{\nu}R^{\lambda}_{\mu}\big)\nonumber\\
&&\hspace{-24pt} +g^{\lambda\tau}\,R_{\mu\nu}+\frac{2}{(m-2)}g_{\mu\nu}\big(R^{\lambda\tau}-\frac{1}{2}R\,g^{\lambda\tau}\big) +(R-2\Lambda)\,\delta^{\lambda\tau}_{\mu\nu},\label{3.4.5}
\end{eqnarray}
with some additional terms that can be found in \cite{HKLT} for the Vilkovisky-DeWitt effective action. 
The inverse of the DeWitt metric is
\begin{equation}
{\cal G}_{\mu\nu\,\rho\sigma}=
\frac{1}{2}(g_{\mu\rho}g_{\nu\sigma}+g_{\mu\sigma}g_{\nu\rho})+\frac{1}{(2-m)}\,g_{\mu\nu}g_{\rho\sigma}.\label{3.4.5b}
\end{equation}
This clearly requires $m\ne2$, and so we do not consider the case $m=2$ for gravity. When expanding the Einstein-Hilbert action to quadratic order in $h_{\mu\nu}$ the term that does not involve any derivatives of $h_{\mu\nu}$ can be written as $\displaystyle{\frac{1}{2}h_{\mu\nu}S^{\mu\nu\lambda\tau}h_{\lambda\tau} }$, and the relation between $S^{\mu\nu\lambda\tau}$ and $\displaystyle{Q_{\mu\nu}{}^{\lambda\tau}}$ in \eqref{3.4.5} is
\begin{equation}
Q_{\mu\nu}{}^{\lambda\tau}= {\cal G}_{\mu\nu\,\alpha\beta}\,S^{\alpha\beta\lambda\tau}.\label{3.4.5c}
\end{equation}
It can be advantageous to write the result for the heat kernel coefficients in terms of $S^{\mu\nu\lambda\tau}$ since it can be read off simply from the expansion of the action; however it is $\displaystyle{Q_{\mu\nu}{}^{\lambda\tau}}$ that enters our basic formalism.

We now require the coefficients that appear when the operator is written in the required form in \eqref{djt1.9}. The results turn out to be
\begin{eqnarray}
\big(A^{\alpha\beta}\big)_{\mu\nu}{}^{\lambda\tau}&=&-\frac{1}{2}\big(\Sigma^{\alpha\beta}\big)_{\mu\nu}{}^{\lambda\tau}-\frac{1}{2}\big(\Sigma^{\beta\alpha}\big)_{\mu\nu}{}^{\lambda\tau} 
,\label{3.4.6}\\
\big(B^{\alpha}\big)_{\mu\nu}{}^{\lambda\tau}&=&\big(\Sigma^{\rho\sigma}\big)_{\mu\nu}{}^{\lambda\tau}\Gamma^{\alpha}_{\rho\sigma}\nonumber\\
&& +\big\lbrack\big(\Sigma^{\alpha\beta}\big)_{\mu\nu}{}^{\rho\sigma} +\big(\Sigma^{\beta\alpha}\big)_{\mu\nu}{}^{\rho\sigma} \big\rbrack\big(\delta^{\tau}_{\sigma}\,\Gamma^{\lambda}_{\beta\rho} + \delta^{\lambda}_{\sigma}\,\Gamma^{\tau}_{\beta\rho} \big),\label{3.4.7}\\
\big(C\big)_{\mu\nu}{}^{\lambda\tau}&=&Q_{\mu\nu}{}^{\lambda\tau} +\big(\Sigma^{\alpha\beta}\big)_{\mu\nu}{}^{\gamma\delta}\big\lbrack \Gamma^{\sigma}_{\beta\gamma,\alpha}\delta^{\lambda\tau}_{\sigma\delta} +\Gamma^{\sigma}_{\beta\delta,\alpha}\delta^{\lambda\tau}_{\sigma\gamma} -\Gamma^{\sigma}_{\alpha\beta}\Gamma^{\rho}_{\sigma\gamma}\delta^{\lambda\tau}_{\rho\delta}\nonumber\\
&& -\Gamma^{\sigma}_{\alpha\beta}\Gamma^{\rho}_{\sigma\delta}\delta^{\lambda\tau}_{\rho\gamma} -\Gamma^{\sigma}_{\alpha\gamma}\Gamma^{\rho}_{\beta\sigma}\delta^{\lambda\tau}_{\rho\delta} -\Gamma^{\sigma}_{\alpha\gamma}\Gamma^{\rho}_{\beta\delta}\delta^{\lambda\tau}_{\rho\sigma} -\Gamma^{\sigma}_{\alpha\delta}\Gamma^{\rho}_{\beta\gamma}\delta^{\lambda\tau}_{\rho\sigma}\nonumber\\
&& -\Gamma^{\sigma}_{\alpha\delta}\Gamma^{\rho}_{\beta\sigma}\delta^{\lambda\tau}_{\rho\gamma}\big\rbrack. \label{3.4.8}
\end{eqnarray}
We now expand these coefficients in Riemann normal coordinates to obtain the expansion coefficients defined in \eqref{djt1.10}--\eqref{djt1.12}. The expansions \eqref{3.1.5}--\eqref{3.1.88} may be used here. Of special interest is (see \eqref{djt1.34} for the definition)
\begin{equation}
\big(P^{\alpha\beta}\big)_{\mu\nu}{}^{\lambda\tau}=\frac{1}{2}\big( \delta^{\alpha\lambda}_{\mu\nu} \delta^{\beta\tau} +\delta^{\alpha\tau}_{\mu\nu} \delta^{\beta\lambda} +\delta^{\beta\lambda}_{\mu\nu} \delta^{\alpha\tau}+\delta^{\beta\tau}_{\mu\nu} \delta^{\alpha\lambda} - 2\,\delta^{\alpha\beta}_{\mu\nu} \delta^{\lambda\tau}\big)\;.\label{3.4.9}
\end{equation}
From \eqref{djt1.35} it can be seen that
\begin{equation}
\big(\hat P\big)_{\mu\nu}{}^{\lambda\tau}= \big( \delta^{\alpha\lambda}_{\mu\nu} \delta^{\beta\tau} +\delta^{\alpha\tau}_{\mu\nu} \delta^{\beta\lambda}  - \delta^{\alpha\beta}_{\mu\nu} \delta^{\lambda\tau}\big)\hat p_{\alpha} \hat p_{\beta},\label{3.4.10}
\end{equation}
which obeys the required relation \eqref{djt1.36}:
\begin{equation}
\big(\hat P\big)_{\mu\nu}{}^{\lambda\tau} \big(\hat P\big)_{\lambda\tau}{}^{\rho\sigma} 
=\,\big(\hat P\big)_{\mu\nu}{}^{\rho\sigma}.\label{3.4.11}
\end{equation}
that was assumed in the derivation of our general results.

The normal coordinate expansions of all of the required coefficients that enter the Green function \eqref{g2} are easily 
found from \eqref{3.1.6} and \eqref{3.1.7}. These are not repeated here for brevity. After following the method described above we find
\begin{eqnarray}\label{e1tensor}
{\rm Tr}\,E_1&=&g^{\mu\nu}g_{\rho\sigma}Q_{\mu\nu}{}^{\rho\sigma}
\left\{\frac{(1+\zeta)^{-m/2}-1}{m}\right\}\nonumber\\
&&+\delta_{\rho\sigma}{}^{\mu\nu}Q_{\mu\nu}{}^{\rho\sigma}
\left\{\frac{2-m-2\,(1+\zeta)^{-m/2}}{m}\right\}\nonumber\\
&&+R\left\{ \frac{(m^3-m^2-12\,m-48)}{12\,m}\label{3.4.13}\right.\nonumber\\
&&+\left.\frac{(1+\zeta)\,m^2+6\,(1+\zeta)\,m+24}{6\,m}\,(1+\zeta)^{-m/2}\right\}.
\end{eqnarray}
(The untraced expression is more lengthy and is not given here or in \ref{appuntraced} for brevity.) As a check on our result, if we let $\zeta\rightarrow0$, so that we have a minimal operator, it can be seen that \eqref{e1tensor} becomes equal to the standard result of ${\rm Tr}(E_1)={\rm Tr}(\frac{1}{6}\,R\,I-Q)$. (For symmetric rank two tensors ${\rm tr}(I)=\frac{1}{2}m(m+1)$.)

In the special case of Einstein gravity, from \eqref{3.4.5}, we find
\begin{eqnarray}
{\rm Tr}\,E_1&=&\Lambda\,\left\lbrack m(m-1)+2m(1+\zeta)^{-m/2}\right\rbrack\\
&&\hspace{-36pt}+R\,\left\lbrace -\,\frac{(5m^2-17m+36)}{12} +\frac{(18-5m+(m+6)\zeta)}{6}\,(1+\zeta)^{-m/2} \right\rbrace\nonumber.
\end{eqnarray}

\section{$E_2$ coefficient}\label{E2}

In this section we will calculate the trace of the $E_2$ coefficient for the three examples that we have discussed previously. As before, we will only look at the trace of the coefficient here for simplicity as this reduces the number of terms involved considerably. It is the traced coefficient that is of direct application to quantum field theory in one-loop calculations, especially in four dimensions where it can be used to describe the renormalisation or cut-off scale dependence of terms in the effective action. The untraced coefficients can be evaluated and will be given elsewhere since they are of interest to quantum field theory calculations beyond one-loop order.

\subsection{Real vector field}\label{real vector E2}

Following theorem \ref{lem1}, can write the traced $E_2$ coefficient in terms of invariants as
\begin{eqnarray}
{\rm Tr}\,E_2&=&a_1\,\nabla^2\,Q+a_2\,Q^{\mu\nu}{}_{;\mu\nu}+ a_3\,\nabla^2\,R + a_4\,Q^2 + a_5\,Q^{\mu\nu}Q_{\mu\nu}+a_6\,RQ\nonumber\\
&&+a_7\,R_{\mu\nu}Q^{\mu\nu}+a_8\,R^2 +a_9\,R_{\mu\nu}R^{\mu\nu}+a_{10}\,R_{\mu\nu\lambda\tau}R^{\mu\nu\lambda\tau}.\label{4.1.1}
\end{eqnarray}
Here $Q=Q^{\mu}{}_{\mu}$, and $Q_{\mu\nu}$ has been assumed to be symmetric. The results for the coefficients 
$a_1\dots  a_{10}$ obtained from the method described in the previous sections are given below.

\begin{eqnarray}
a_1&=& \frac{(-m^4 \zeta^2+5 m^3 \zeta^2-2 m^2 \zeta^2-32 m \zeta^2+96 \zeta^2+192 \zeta+96)}{6 (m-4) (m^2-4) m  \zeta^2}\nonumber\\
&&- (1+\zeta)^{1-m/2}\frac{ \left(m^3 \zeta^2+6 m^2 \zeta^2+8 m \zeta^2+48 m \zeta+96 \zeta+96\right)}{6 (m-4) (m^2-4) m  \zeta^2}
,\label{4.1.2a}\\
a_2&=&-\left( 1+\zeta \right) ^{1-m/
2}{\frac { \left( m^3\zeta^2-4m\zeta^2-24m\zeta-48m-48\zeta \right)  }{3\, \left( m-4 \right) m\zeta^2 \left( {m}^{2}-4 \right) }}\nonumber\\
&&\hspace{-36pt}-{
\frac {4\,\zeta^2m+5\,{m}^{3}\zeta^2-24\,{m}^{2}\zeta^2+48\,\zeta^2+72
\,m\zeta+48\,\zeta-24\,{m}^{2}\zeta+48\,m}{3\, \left( m-4 \right) m\zeta^2 \left( {m}
^{2}-4 \right) }}
,\label{4.1.2b}\\
a_3&=&\frac{1}{30 (m-4) (m^2-4) m  \zeta^2} \Big\lbrack 240(m-2) -120 (6-3m+m^2)\zeta\nonumber\\
&&\qquad +\left(m^5-5 m^4+15 m^3-70 m^2+104 m-240\right) \zeta^2\Big\rbrack\nonumber\\
&&+\frac{(1+\zeta)^{1-\frac{m}{2}}}{30 (m-4) (m^2-4) m  \zeta^2}\Big\lbrack m(m^2-4) (m+1) \zeta^3\nonumber\\
&& +m (m+1) (m+2) (m+8) \zeta^2+120 (m+2) \zeta-240(m-2)\Big\rbrack,\label{4.1.2c}
 \\
a_4&=&{\frac { \left( 4+2\zeta+ m\zeta \right)  }{2\,m \left( {m}^{2}-4 \right)\zeta }}\left( 1+\zeta \right) ^{-m/2}+{\frac { \left( m\zeta
-2\zeta-4 \right) }{2\,m \left( {m}^{2}-4 \right)\zeta }},\label{4.1.2d}\\
a_5&=&-{\frac { \left( 4\,\zeta+4\,m+2 m\zeta \right)  }{2\,m \left( {m}^{2}-4 \right)\zeta }}\left( 1+\zeta
 \right) ^{-m/2}\nonumber\\
 &&\qquad-{\frac { \left( 2 m\zeta-4\,m-4\zeta-{m}^{3}\zeta+2\,{m}^{2}\zeta \right) }{2m \left( {m
}^{2}-4 \right)\zeta }},\label{4.1.2e}\\
a_6&=&{\frac { \left( -{m}^{2}\zeta^2-2\,m\zeta^2-{m}^{2}\zeta-8m\zeta-12\zeta-24 \right) }{6m \left( {m}^{2}-4
 \right)\zeta }} \left( 1+\zeta \right) ^{-m/2}\nonumber\\
 &&+\,{\frac { \left( -{m}^{3}\zeta+{m}^{2}\zeta+12\zeta+24
 \right) }{6m \left( {m}^{2}-4 \right)\zeta }},\label{4.1.2f}\\
 a_7&=&{\frac {\left( {m}^{2}\zeta^2+2m\zeta^2+8m\zeta+{m}^{2}\zeta
+12\zeta+12\,m \right)  }{3m \left( {m}^{2}
-4 \right)\zeta }}\left( 1+\zeta \right) ^{-m/2}\nonumber\\
&&-\,{\frac { \left( 12\zeta+8m\zeta-5\,{m}^{2}\zeta+12\,
m \right) }{3m \left( {m}^{2}-4 \right)\zeta }},\label{4.1.2g}\\
a_8&=& \frac{\left(m^4-m^3-16 m^2+16 m-72\right) \zeta-144}{72 (m^2-4) m  \zeta}\nonumber\\
&&+\frac{(1+\zeta)^{-m/2}}{{72 (m^2-4) m  \zeta}}\Big\lbrack 144 +(m+2) (m^2+10m+36)\zeta\nonumber\\
&&\qquad +2 m (m+2) (m+4)\zeta^2+m \left(m^2-4\right)\zeta^3 \Big\rbrack,\label{4.1.2h}\\
a_9&=&\frac{360 m+\left(-m^4+m^3-116 m^2+296 m+360\right) \zeta}{180 (m^2-4) m \zeta}\label{4.1.2i}\\
&&-\frac{(1+\zeta)^{-m/2}}{{180 (m^2-4) m \zeta}}\Big\lbrack 360m +(m+2) (m^2+58m+180)\zeta\nonumber\\
&&\qquad+2 m \left(m^2+30 m+56\right)\zeta^2+m(m^2-4)\zeta^3 \Big\rbrace,\nonumber\\
 a_{10}&=& \frac{1}{180} \left( 1+\zeta \right) 
^{2-m/2}+{\frac {1}{180}}\, \left( m-16 \right).\label{4.1.2j}
\end{eqnarray}

\begin{table}[htb]
\caption{\label{table4}Limiting form of the sub-coefficients in the traced heat kernel coefficient ${\rm Tr}\,E_2$ for 
vectors for minimal operators $\zeta=0$ or spacetime dimension $m=4$. }
\centering
\begingroup
\renewcommand{\arraystretch}{1.5}
\begin{tabular}{>{$}l<{$}>{$\displaystyle}c<{$}>{$\displaystyle}l <{$}}
{\rm Term}&\zeta=0&m=4\\
\hline\noalign{\smallskip}
a_1&-\frac{1}{6}&{\frac { \left( 1+2\zeta \right) \ln  \left( 1+\zeta \right) }{6\,\zeta^2
}}-\,{\frac {8\,\zeta^2+21\,\zeta+6}{36\,\zeta \left( 1+\zeta \right) }}\label{4.1.3a}\\
a_2&0&-{\frac { \left( 4-\zeta \right) \ln  \left( 1+\zeta \right) }{6\,\zeta^2}}-
{\frac {13\,\zeta^2-6\,\zeta-24}{36\,\zeta \left( 1+\zeta \right) }}\label{4.1.3b}\\
a_3&\frac{m}{30}&-{\frac { \left( \zeta^2+5\,\zeta-2 \right) \ln  \left( 1+\zeta \right) 
}{12\,\zeta^2}}+{\frac {133\,\zeta^2+168\,\zeta-60}{360\,\zeta
 \left( 1+\zeta \right) }}\label{4.1.3c}\\
 a_4&0&{\frac {\zeta^2}{ 48\,\left( 1+\zeta \right) ^{2}}}\label{4.1.3d}\\
 a_5&\frac{1}{2}&{\frac {12+18\,\zeta+7\,\zeta^2}{24\, \left( 1+\zeta \right) ^{2}}}\label{4.1.3e}\\
a_6&-\frac{1}{6}&-\,{\frac {4+6\,\zeta+3\,\zeta^2}{24\, \left( 1+\zeta \right) ^{2}}}\label{4.1.3f}\\
a_7&0&{\frac {\zeta \left( 4+3\,\zeta \right) }{12\, \left( 1+\zeta \right) ^{2}}}\label{4.1.3g}\\
a_8&\frac{m}{72}&-\,{\frac {\zeta^2-4\,\zeta-8}{144\, \left( 1+\zeta \right) ^{2}} }\\
a_9&-\,\frac{m}{180}&-\,{\frac {23\,\zeta^2+46\,\zeta+8}{360\, \left( 1+\zeta \right) ^{2}}}\label{4.1.3i}\\
a_{10}&\frac{m-15}{180}&-\frac{11}{180}\label{4.1.3j}\\
\noalign{\smallskip}\hline
\end{tabular}
\endgroup
\end{table}

The limits of our results for $\zeta=0$ and for $m=4$ are given in table~\ref{table4}. These results are in agreement with \cite{Branson:1999jz} for the terms that do not involve derivatives. In the limit as $m\rightarrow4$ the results of \cite{barvinsky1985generalized} are recovered, again for the terms that do not involve derivatives. For $\zeta=0$ we recover the results of Gilkey~\cite{gilkey1975spectral}.

\subsection{Yang-Mills field}\label{Yang-Mills E2}

The operator which we use for the Yang-Mills field was described in Sect. \ref{yang mills}. Another application of theorem \ref{lem1}
allows us to write
\begin{eqnarray}
{\rm tr}\,E_2&=&a_1\,\nabla^2{Q}^{a \beta}{}_{a \beta}+ \frac{1}{2}\,a_2\,\big({Q}^{a \alpha}{}_{ a}{}^{ \beta}{}_{;\beta\alpha} +{Q}^{a \beta}{}_{ a}{}^{ \alpha}{}_{;\beta\alpha} \big)+ d_G\,a_3\,\nabla^2\,R\nonumber\\
&& + a_4\,{Q}^{a \alpha}{}_{ b \alpha} {Q}^{b \beta}{}_{ a \beta}  
+ a^{(1)}_{5}\,{Q}^{a \alpha}{}_{ b}{}^{ \beta} {Q}^{b}{}_{ \alpha}{}_{a \beta}
+a^{(2)}_{5}\,{Q}^{a \alpha}{}_{ b}{}^{ \beta} {Q}^{b}{}_{ \beta}{}_{a \alpha} +a_6\,RQ^{a\alpha}{}_{a\alpha}\nonumber\\
&&+a_7\,R_{\mu\nu}Q^{a\mu}{}_{a}{}^{\nu}+d_G \,a_8\,R^2 +d_G\,a_9\,R_{\mu\nu}R^{\mu\nu}+d_G\,a_{10}\,R_{\mu\nu\lambda\tau}R^{\mu\nu\lambda\tau}\nonumber\\
&&+a_{11}\,g\,f_{a b c}\,F^{a}{}_{\mu\nu}\,Q^{b\mu c \nu}+ a_{12}\,g^2\,C_2\,F^{a}{}_{\mu\nu}\,F^{a\mu\nu},\label{4.2.1}
\end{eqnarray}
where $C_2$ is the quadratic Casimir invariant of the adjoint representation of the gauge group.
The results for the coefficients $a_1,\ldots,a_4$ and $a_6,\ldots,a_{10}$ are the same as those for the real vector field given in \eqref{4.1.2a}--\eqref{4.1.2d}, and \eqref{4.1.2f}--\eqref{4.1.2j}. The new coefficients are given below. The limiting forms of these extra coefficients for $\zeta=0 $ or $m=4$ are given in table~\ref{table5}.
\begin{eqnarray}
a^{(1)}_{5}&=&\frac{m\zeta-2\zeta-4}{2 m \left(m^2-4\right) \zeta}\nonumber\\
&&+(1+\zeta)^{-m/2}\frac{ (m \zeta+2\zeta+4)}{2 m \left(m^2-4\right) \zeta}
,\label{4.2.2a}\\
a^{(2)}_{5}&=&\frac{m^3 \zeta-2 m^2 \zeta-3 m \zeta+4 m+6 \zeta+4}{2 m \left(m^2-4\right) \zeta}\nonumber\\
&&-(1+\zeta)^{-m/2}\frac{ (3 m \zeta+4 m+6 \zeta+4)}{2 m \left(m^2-4\right) \zeta}
,\label{4.2.2b}\\
a_{11}&=&\frac{3 m \zeta-8 \zeta-8}{(m-2) m \zeta}+ (1+\zeta)^{1-m/2}\frac{ (m \zeta+8)}{(m-2) m \zeta}
,\label{4.2.2c}\\
a_{12}&=&\frac{(-m^3 \zeta+3 m^2 \zeta-26 m \zeta+96 \zeta+96)}{12 (m-2) m \zeta}\nonumber\\
&&+ (1+\zeta)^{1-m/2}\frac{ \left(-m^2 \zeta^2-m^2 \zeta+2 m \zeta^2-22 m \zeta-96\right)}{12 (m-2) m \zeta}
.\label{4.2.2d}
\end{eqnarray}

\begin{table}[htb]
\caption{\label{table5}Limiting form of new sub-coefficients in the traced heat kernel coefficient ${\rm Tr}\,E_2$ for 
 Yang-Mills fields with minimal operators $\zeta=0$ or spacetime dimension $m=4$. }
\centering
\begingroup
\renewcommand{\arraystretch}{1.5}
\begin{tabular}{>{$}l<{$}>{$\displaystyle}c<{$}>{$\displaystyle}c <{$}}
\hline\noalign{\smallskip}
{\rm Term}&\zeta=0&m=4\\
\noalign{\smallskip}\hline\noalign{\smallskip}
a^{(1)}_{5}&0&{\frac {\zeta^2}{48\,(1+\zeta)^{2}}}
\label{4.2.3a}\\
a^{(2)}_{5}&\frac12&{\frac {(13\,\zeta^2+24+36\,\zeta)}{48\,(1+\zeta)^{2}}}
\label{4.2.3b}\\
a_{11}&0&{\frac {\zeta}{2\,(1+\zeta)}}
\label{4.2.3c}\\
a_{12}&-\frac{m}{12}&-\frac{1}{3}\\
\noalign{\smallskip}\hline
\end{tabular}
\endgroup
\end{table}

In the case of pure Yang-Mills, we have
\begin{equation}
Q^{a\alpha}{}_{b\beta}=\delta^{a}_{b}\,R^{\alpha}{}_{\beta}+2\,g\,f^{a}{}_{ b c}\,F^{c\,\alpha}{}_{\beta}.\label{4.2.6}
\end{equation}
The number of invariants is reduced to just five, with the traced heat kernel coefficient now 
\begin{eqnarray}
{\rm tr}\,E_2&=&d_G\,a_3^\prime\,\nabla^2 R + d_G\,a_8^\prime R^2 + d_G\,a_9^\prime\,R^{\mu\nu}R_{\mu\nu}\nonumber\\
&& + d_G\,a_{10}^\prime\,R^{\mu\nu \lambda \tau}R_{\mu\nu\lambda\tau} +a_{12}^\prime g^2\,C_2(G_{\rm adj})\,F^{a}{}_{\mu\nu}F^{a\mu\nu},\label{4.2.7}
\end{eqnarray}
The new coeficients are related to the old ones by
\begin{eqnarray}
a_3^\prime&=&a_1+\frac{1}{2}\,a_2+a_3,\\
a_8^\prime&=&a_4+a_6+a_8,\\
a_9^\prime&=&a_5^{(1)}+a_{5}^{(2)}+a_7+a_9,\\
a_{10}^\prime&=&a_{10}.
\end{eqnarray}
After substituting the earlier coefficients we find,
\begin{eqnarray}
a_3^\prime
&=& \frac{m^2-10 m+19}{30 (m-4)}+\frac{(m+1)  (1+\zeta)^{2-m/2}}{30 (m-4)},\label{4.2.8a}\\
a_8^\prime
&=& \frac{1}{72} (1+\zeta)^{2-m/2}+\frac{m-13}{72},\label{4.2.8b}\\
a_9^\prime
&=&\frac{91-m}{180}-\frac{1}{180} (1+\zeta)^{2-m/2}.\label{4.2.8c}\\
a_{12}^\prime&=&\frac{25-m}{12}-\frac{1}{12}  (1+\zeta)^{2-m/2}.\label{4.2.8d}
\end{eqnarray}

\begin{table}[htb]
\caption{\label{table6}Limiting form of coefficients in the traced heat kernel coefficient ${\rm Tr}\,E_2$ for 
pure Yang-Mills fields with minimal operators $\zeta=0$ or spacetime dimension $m=4$. }
\centering
\begingroup
\renewcommand{\arraystretch}{1.5}
\begin{tabular}{>{$}l<{$}>{$\displaystyle}c<{$}>{$\displaystyle}c <{$}}
\hline\noalign{\smallskip}
{\rm Term}&\zeta=0&m=4\\
\noalign{\smallskip}\hline\noalign{\smallskip}
a_3^\prime&\frac{m}{30}-\frac16&-\frac{ 5 \log (1+\zeta)+2}{60}\\
a_8^\prime&\frac{m}{72}-\frac16&-\frac{1}{9}\\
a_9^\prime&\frac{1}{2}-\frac{m}{180}&\frac{43}{90}\\
a_{10}^\prime&\frac{m}{180}-\frac{1}{12}&-\frac{11}{180}\\
a_{12}^\prime&2-\frac{m}{12}&\frac{5}{3}.\label{4.2.9e}\\
\noalign{\smallskip}\hline
\end{tabular}
\endgroup
\end{table}
The limiting forms of the coefficients for minimal operators or for $m=4$ are given in table \ref{table6}.
The most important coefficient here is $a^\prime_{12}$, which gives the pure Yang-Mills contribution to the
renormalisation group  $\beta-$function of the gauge coupling $g$,
\begin{equation}
\beta(g)=-\frac{g^3}{ 8\pi^2}a^\prime_{12}C_2.
\end{equation}
The full Yang-Mills result for this $\beta-$function can be obtained by adding a ghost field contribution of $1/6$
to $a_{12}^\prime$.
Perhaps the most remarkable feature of the table is that only the coefficient of the $\nabla^2 R$ term has a dependence 
on the non-minimal parameter $\zeta$ in four dimensions. Ordinarily, each term in the integrated heat kernel coefficient
defines a generalised type of $\beta-$function by adding Vilkovisky-DeWitt corrections or taking the limit $\zeta\to\infty$. 
The pure Yang-Mills case is special, in the sense that the Vilkovisky corrections to these $\beta-$functions vanish.

In the case of $m=4$ the results for the terms that are not total derivatives agree with Barvinsky and Vilkovisky \cite{barvinsky1985generalized}. Most of the terms here for general $m$ agree with those of Gusynin and Kornyak \cite{gusynin1997computation} and \cite{gusynin1999complete}, although there are some minor differences and the results of these two references are not in complete agreement with each other. Our results do agree with some of the expressions in \cite{gusynin1997computation} and some of those in \cite{gusynin1999complete} so that the most likely explanation is minor typographical errors in these two references.  

\subsection{Gravity}\label{gravity E2}
The operator for rank two tensors was described in Sect. \ref{gravity}. 
Application of theorem \ref{lem1} allows us to express the trace of $E_2$ in the form
\begin{eqnarray}
{\rm Tr}\,E_{2}&=&t_{1}\,S^{\alpha \beta \mu}{}_{\mu;\alpha \beta} +\frac{1}{6(m-2)}\,\nabla^2 S^{\alpha}{}_{\alpha}{}^{\beta}{}_{\beta} + t_{2}\,S^{\alpha\mu\beta}{}_{\mu;\alpha\beta}+ t_{3}\,\nabla^2 S^{\alpha\mu}{}_{\alpha\mu}\nonumber\\
&&+ t_{4}\,S^{\alpha}{}_{\alpha}{}^{\beta\mu}S^{\nu}{}_{\nu\beta\mu} + \frac{1}{2(m-2)^2}\,\big(S^{\alpha}{}_{\alpha}{}^{\beta}{}_{\beta}\big)^2\nonumber\\
&&+t_{5}\,S_{\alpha\beta\mu\nu} S^{\alpha\mu\beta\nu} +t_{5}\,S^{\alpha\beta}{}_{\alpha\mu}S_{\beta\nu}{}^{\mu\nu} +t_{6}\,S_{\alpha\beta\mu\nu} S^{\alpha\beta\mu\nu}\nonumber \\
&&+ \frac{1}{6(m-2)}\,RS^{\alpha}{}_{\alpha}{}^{\beta}{}_{\beta}+t_{7}\,RS^{\alpha\beta}{}_{\alpha\beta}+t_{8}\,R_{\alpha\beta}S^{\alpha\beta\mu}{}_{\mu}\nonumber\\
&&+t_{9}\,R_{\alpha\beta}S^{\alpha\mu\beta}{}_{\mu}+t_{10}\,R_{\alpha\beta\mu\nu}S^{\alpha\mu\beta\nu} +t_{11}\,\nabla^2 R+t_{12}\,R^2\nonumber\\
&&+t_{13}\,R_{\mu\nu}R^{\mu\nu}+ t_{14}\,R_{\mu\nu\alpha\beta}R^{\mu\nu\alpha\beta}\;.\label{4.3.1}
\end{eqnarray}
The coefficients $t_{11},\ldots,t_{14}$ that follow after applying the general formalism are,
\begin{eqnarray}
t_{1}&=&-\,\frac{4 \left((1+\zeta) (4+m\zeta) (1+\zeta)^{-m/2}+m \zeta-4 \zeta-4\right)}{(m-4) (m-2) m \zeta},\label{4.3.2a}\\
t_{2}&=&-\,\frac{2 (1+\zeta) \left((m-8) m (m+2) \zeta^2-48 (m+2) \zeta-48 m\right)}{3 m (m-4) (m^2-4) \zeta^2} \nonumber\\
&&-\,\frac{2 (1+\zeta)^{m/2} }{3 m (m-4) (m^2-4) \zeta^2}\Big\lbrack (m-4) (5m^2-10m-24) \zeta^2\nonumber\\
&&\qquad-24 (m^2-4m-4) \zeta+48 m\Big\rbrack
,\label{4.3.2b}\\
t_{3}&=&\frac{(-m^4\zeta^2+6 m^3 \zeta^2-8 m^2 \zeta^2-48 m \zeta^2+192 \zeta^2+384 \zeta+192}{6 m (m-4) (m^2-4) \zeta^2} \nonumber\\
&&- 2 (1+\zeta)^{1-{m}/{2}}\,\frac{m \left(m^2+6 m+8\right) \zeta^2+48 (m+2) \zeta+96}{6 m (m-4) (m^2-4) \zeta^2},
\label{4.3.2c}\\
t_{4}&=&-\frac{m^2 \zeta-4 m \zeta+4 \zeta+4}{(m-2)^2 m \zeta}+4\, (1+\zeta)^{1-m/2}\frac{1}{(m-2)^2 m \zeta},\label{4.3.2d}\\
t_{5}&=&\frac{2 (m \zeta+2 \zeta+4) (1+\zeta)^{-\frac{m}{2}}+2 (m \zeta-2 \zeta-4)}{m \left(m^2-4\right) \zeta },\label{4.3.2e}\\
t_{6}&=&\frac{m^3 \zeta-4 m^2 \zeta+8 m+8 \zeta+(-4 m \zeta-8 m-8 \zeta) (1+\zeta)^{-\frac{m}{2}}}{2 m \left(m^2-4\right) \zeta }, \label{4.3.2f}\\
t_{7}&=&\frac{- m^3 \zeta+2 m^2 \zeta-4 m \zeta+24 \zeta+48}{6 m \left(m^2-4\right) \zeta }\nonumber\\
&&-(1+\zeta)^{-m/2}\,\frac{(2 m^2 \zeta^2+2 m^2 \zeta+4 m \zeta^2+16 m \zeta+24 \zeta+48) } {6 m \left(m^2-4\right) \zeta},\label{4.3.2g}\\
t_{8}&=&\frac{(-4 m \zeta-8 \zeta-16) (1+\zeta)^{-\frac{m}{2}}-4 m \zeta+8 \zeta+16}{m \left(m^2-4\right) \zeta} ,\label{4.3.2h}\\
t_{9}&=&\frac{2 \left(5 m^2 \zeta-20 m \zeta-12 m+36 \zeta+72\right)}{3 m \left(m^2-4\right) \zeta }\nonumber\\
&&\hspace{-12pt}+2\,(1+\zeta)^{-m/2}\,\frac{(-2 m^2 \zeta^2+m^2 \zeta-4 m \zeta^2-16 m \zeta+12 m-36 \zeta-72)}{3 m \left(m^2-4\right) \zeta },
\label{4.3.2i}\\
t_{10}&=&\frac{-6 m^2 \zeta+8 m \zeta+16 m+24 \zeta+16}{m \left(m^2-4\right) \zeta }\nonumber\\
&&-\,(1+\zeta)^{-m/2}\,\frac{(2 m^2 \zeta^2+2 m^2 \zeta+4 m \zeta^2+16 m \zeta+16 m+24 \zeta+16)}{m \left(m^2-4\right) \zeta}, \label{4.3.2j}\\
t_{11}&=&\frac{1}{60m (m-4) (m+2) \zeta^2}\Big\lbrack(m^5-3 m^4+24 m^3-172 m^2-480 m+960) \zeta^2\nonumber\\
&&+240(m^2-12) \zeta+480(m+4)\Big\rbrack\nonumber\\
&&\hspace{-36pt}+\frac{(1+\zeta)^{-m/2}}{60m (m-4) (m+2) \zeta^2}\,\Big\lbrack 2 m^2(m+2)(m+6)\zeta^4+2m^2(m+2)(2m+17)\zeta^3\nonumber\\
&&\hspace{-24pt}+2(m-4)(m+2)(m^2+15m+60)\zeta^2-960(m+3)\zeta-480(m+4)\Big\rbrack, \label{4.3.2k}\\
t_{12}&=&\frac{m^5 \zeta-m^4 \zeta-28 m^3 \zeta-68 m^2 \zeta+96 m \zeta-2016 \zeta-288 m-2880}{144 m \left(m^2-4\right) \zeta} \nonumber\\
&&\hspace{-36pt}+\frac{(1+\zeta)^{-m/2}}{144 m \left(m^2-4\right) \zeta}\,\Big\lbrack 2m(m^2-4)(m+12)\zeta^3+4m(m+2)(m^2+10m+48)\zeta^2\nonumber\\
&&\quad+2(m+2)(m^3+10m^2+84m+504)\zeta+288(m+10)\Big\rbrack, \label{4.3.2l}\\
\hspace{-24pt}t_{13}&\hspace{-24pt}=&\hspace{-24pt}\frac{(- m^5  +m^4-236 m^3 -4 m^2+3120 m-10080 )\zeta+720 m^2 +1440 m -14400}{360m (m-2)(m+2) \zeta} \nonumber\\
&&+\frac{(1+\zeta)^{-m/2}}{180m (m-2)(m+2) \zeta}\,\Big\lbrack -m(m-2)(m+2)(m-90)\zeta^3\nonumber\\
&&\hspace{-36pt}-2m(m+2)(m^2+28m-240)\zeta^2 -(m+2)(m^3+58m^2+240m-2520)\zeta\nonumber\\
&&-360(m^2+2m-20)\Big\rbrack, \label{4.3.2m}\\
t_{14}&=&\frac{(m^5 -31 m^4 -34 m^3 -956 m^2 +3360 m +6480)\zeta+4320( m+1)}{360 m \left(m^2-4\right) \zeta} \nonumber\\
&&+\frac{(1+\zeta)^{-m/2}}{360 m \left(m^2-4\right) \zeta}  \Big\lbrack 2m(m^2-4)(m-15)\zeta^3\nonumber\\
&&\quad+4m(m+2)(m^2-17m-240)\zeta^2\nonumber\\
&&\quad+2(m+2)(m^3-17m^2-510m-1620)\zeta-4320(m+1)\Big\rbrack , \label{4.3.2n}
\end{eqnarray}

As a check on the results, in the minimal case where $\zeta=0$ we have the results given in the first column of table~\ref{table7}. 
These are in exact agreement with what is found from using the standard formula \cite{gilkey1995invariance}. 
The $m=4$ results are given in the second column. Note that the coefficients which diverge in the $\zeta\to\infty$
limit are all terms that multiply total derivatives.

\begin{table}[htb]
\caption{\label{table7}Limiting form of coefficients in the traced heat kernel coefficient ${\rm Tr}\,E_2$ for 
the gravitational field with minimal operators $\zeta=0$ or spacetime dimension $m=4$. }
\centering
\begingroup
\renewcommand{\arraystretch}{1.2}
\begin{tabular}{>{$}l<{$}>{$\displaystyle}c<{$}>{$\displaystyle}c <{$}}
\hline\noalign{\smallskip}
{\rm Term}&\zeta=0&m=4\\
\noalign{\smallskip}\hline\noalign{\smallskip}
t_{1}&0&{-\frac{(\zeta+2) \zeta-2 (1+\zeta) \log (1+\zeta)}{2 (1+\zeta) \zeta}}\\ 
t_{2}&0&-\frac{2 \left(\zeta \left(\zeta^2-6 \zeta-6\right)+3 \left(\zeta^2+3 \zeta+2\right) 
\log (1+\zeta)\right)}{9 (1+\zeta) \zeta^2}\\ 
t_{3}&-1/6&{\frac{\zeta \left(-5 \zeta^2-18 \zeta-6\right)+6 \left(2 \zeta^2+3 \zeta+1\right) 
\log (1+\zeta)}{18 (1+\zeta) \zeta^2}}\\ 
t_{4}&-1/(m-2)&{-\frac{\zeta+2}{4 (1+\zeta)}}\\ 
t_{5}&0&{\frac{\zeta^2}{12 (1+\zeta)^2}}\\ 
t_{6}&1/2&{\frac{\zeta^2+6 \zeta+6}{12 (1+\zeta)^2}}\\ 
t_{7}&-1/6&{-\frac{\zeta^2+2 \zeta+2}{12 (1+\zeta)^2}}\\ 
t_{8}&0&{-\frac{\zeta^2}{6 (1+\zeta)^2}}\\ 
t_{9}&0&{\frac{\zeta (3 \zeta+4)}{6 (1+\zeta)^2}}\\ 
t_{10}&0&-{\frac{(6+5 \zeta) \zeta}{6 (1+\zeta)^2}}\\ 
t_{11}&m(m+1)/60&{\frac{4 \zeta \left(2 \zeta^2-3\right)-\left(6 \zeta^3+9 \zeta^2-9 \zeta-12\right) 
\log (1+\zeta)}{9 (1+\zeta) \zeta^2}}\\ 
t_{12}&m(m+1)/144&-{\frac{23 \zeta^2+16 \zeta-10}{72 (1+\zeta)^2}}\\ 
t_{13}&-m(m+1)/360&-{\frac{11 \zeta^2+28 \zeta+2}{36 (1+\zeta)^2}}\\ 
t_{14}&(m^2-29 m-60)/360&-{\frac{23 \zeta^2+64 \zeta+32}{72 (1+\zeta)^2}}\\ 
\noalign{\smallskip}\hline
\end{tabular}
\endgroup
\end{table}

In the case of Einstein gravity, with a cosmological constant $\Lambda$, we have six invariants,
\begin{eqnarray}
{\rm tr}\,E_2&=&e_1\,\nabla^2 R +e_2\,R_{\alpha\beta \gamma \delta} R^{\alpha\beta\gamma\delta} + e_3\, R_{\alpha\beta}R^{\alpha\beta}\nonumber\\
&+&e_4\,R^2+e_5\,R\Lambda+ e_6\,\Lambda^2.\label{4.3.3}
\end{eqnarray}
After substituting for $S^{\mu\nu\rho\sigma}$ in \eqref{4.3.1}, the coefficients are as follows:
\begin{eqnarray}
e_{1}&=&\frac{(-2 m^4 +11 m^3 -28 m^2 -116 m +480)\zeta^2+960 \zeta+480}{30 (m-4) (m+2) \zeta^2}\nonumber\\
&&+\frac{(1+\zeta)^{-m/2}}{30 (m-4) (m+2) \zeta^2} \Big\lbrack m(m+2)(m+6)\zeta^4-m(m+2)(3m+8)\zeta^3\nonumber\\
&&\quad-2(m+2)(2m^2+7m+120)\zeta^2 -240(m+4)\zeta-480\Big\rbrack\label{4.3.4a}\\
e_{2}&=&\frac{(m^2-31 m+510)}{360}+\frac{(m-15)}{180} \,(1+\zeta)^{2-m/2}\label{4.3.4b}\\
e_{3}&=&\frac{1}{360 (m-2)^2 (m+2) \zeta}\Big\lbrack1440 \left(3 m^2-6 m-8\right)
\nonumber\\
&&\quad-(m-2) \left(m^4-181 m^3+3176 m^2-2636 m-13680\right) \zeta\Big\rbrack\nonumber\\
&&+\frac{(1+\zeta)^{-m/2}}{180 (m-2)^2 (m+2) \zeta}\Big \lbrack - (m-90) (m-2)^2 (m+2)\zeta^3 \nonumber\\ 
&&\quad -2 (m^2-4)  ( m^2-32m+540)\zeta^2\nonumber\\
&&\quad - (m+2) \left(m^3+26 m^2+2284 m-6120\right)\zeta\nonumber\\
&&\quad-720 \left(3 m^2-6 m-8\right) \Big \rbrack\label{4.3.4c}\\
e_{4}&=&\frac{1}{144m (m-2)^2 (m+2) \zeta}\Big\lbrack -576(3m^2-6m-8)\nonumber\\
&&\quad (25 m^6-195 m^5+622 m^4+396 m^3-5192 m^2+3480 m+4608)\zeta  \Big\rbrack\nonumber\\
&&+\frac{(1+\zeta)^{-m/2}}{72m (m-2)^2 (m+2) \zeta} \Big\lbrack  (m-2)^2 m (m+2) (m+12)\zeta^3 \nonumber\\&&\quad -2m (m^2-4)  \left(5 m^2+2 m-120\right)\zeta^2\nonumber\\
&&\quad+  (2 + m) ( 25 m^4- 280 m^3 +1156 m^2- 1104 m-1152)\zeta\nonumber\\
&&\quad+288 \left(3 m^2-6 m-8\right)  \Big \rbrack\label{4.3.4d}\\
e_{5}&=&\frac{1}{6} (-5 m^2+17 m-36)
+\frac{1}{3} \left\lbrack (m+6)\zeta+18-5m \right\rbrack (1+\zeta)^{-m/2}\label{4.3.4e}\\
e_{6}&=&m(m-1)+ 2m\, (1+\zeta)^{-m/2}\label{4.3.4f}
\end{eqnarray}

\begin{table}[htb]
\caption{\label{table8}Limiting form of coefficients in the traced heat kernel coefficient ${\rm Tr}\,E_2$ for 
Einstein gravity with minimal operators $\zeta=0$ or spacetime dimension $m=4$. }
\centering
\begingroup
\renewcommand{\arraystretch}{1.3}
\begin{tabular}{>{$}l<{$}>{$\displaystyle}c<{$}>{$\displaystyle}c <{$}}
\hline\noalign{\smallskip}
{\rm Term}&\zeta=0&m=4\\
\noalign{\smallskip}\hline\noalign{\smallskip}
e_{1}&\frac{1}{30} (3-2 m) m&-\,\frac{2\zeta(5\zeta^2+18\zeta+16)+6(1+\zeta)(\zeta^2-4\zeta-2)
 \log (1+\zeta)}{9 (1+\zeta) \zeta^2}\label{4.3.5a}\\
e_{2}&\frac{1}{360} \left(m^2-29 m+480\right) &\frac{19}{18} \label{4.3.5b}\\
e_{3}&-\frac{m^3-181 m^2+1438 m-720}{360 (m-2)}&-\frac{55 \zeta^2+122 \zeta+55}{18 (1+\zeta)^2} \label{4.3.5c}\\
e_{4}&\frac{25 m^3-145 m^2+262 m+144}{144 (m-2)} &\frac{71 \zeta^2+118 \zeta+59}{36 (1+\zeta)^2} \label{4.3.5d}\\
e_{5}&\frac{1}{6} (7-5 m) m&-\frac{2 \left(12 \zeta^2+19 \zeta+13\right)}{3 (1+\zeta)^2} \label{4.3.5e}\\
e_{6}&m (m+1)&\frac{4 \left(3 \zeta^2+6 \zeta+5\right)}{(1+\zeta)^2} \label{4.3.5f}\\
\noalign{\smallskip}\hline
\end{tabular}
\endgroup
\end{table}

The limiting form of the coefficients for minimal operators (with $\zeta=0$) or spacetime dimension four ($m=4$) are given in table
\ref{table8}. The results for a minimal operator with $m=4$ were first worked out by 
Christensen and Duff \cite{duff1980} for gravity with a cosmological constant. The table agrees with their results in the case they considered which had $R_{\mu\nu}=\Lambda g_{\mu\nu}$.

Unlike in the pure Yang-Mills case, the coefficients for $m=4$ depend on the parameter $\zeta$
and in general they have a different value from case of a minimal operator.  In order to define $\beta-$functions
(in the generalised sense since the terms are non-renormalizable) we have to add in the ghost field contributions
and take the Vilkovisky-DeWitt limit $\zeta\to\infty$. The minimal operator, without Vilkovisky-DeWitt corrections, 
does not give the correct $\beta-$functions.

\section{Discussion}
We have shown that, for the special class of operators which are important in
quantum field theory, the heat kernel coefficients belong to the finite algebra of invariants at the given order,
just as they do for the minimal operator.  The coefficients of the individual terms depend on the spacetime 
dimension and on the parameter $\zeta$ which controls the non-minimal term. Many of these coefficients
are tabulated in the text, but we realise that they are unwieldy and so computer files containing the 
coefficients can be obtained by contacting the authors.

The reliability of the calculations is a crucial issue. In order to reduce errors we have used two different 
methods to evaluate the heat kernel coefficient $E_1$, both based on momentum space expansions.
The first method gave a general formula for the trace ${\rm Tr}\,E_1$, and the second method used 
a {\tt Cadabra} program to obtain results for vector, Yang-Mills and tensor fields. The agreement between
the two methods gives us some confidence that the {\tt Cadabra} program should also give reliable results
for ${\rm Tr}\,E_2$. We have checked our results against all the previously known results for the heat
kernel ceofficients of non-minimal operators.

An important application of non-minimal operators is in the evaluation of the Vilkovisky-DeWitt effective action.
The heat kernel coefficients tell us how terms in the action vary with changes in the cut-off scale, 
and can be used to define $\beta-$functions. The complexity of the non-minimal heat kernel coefficients,
combined with the fact that the Vilkovisky corrections vanish for Yang-Mills theory, 
has lead to the widespread use of minimal operators throughout quantum field theory.
However, for tensor fields like gravity, the dependence of the heat kernel on the parameter
$\zeta$ indicates that we have to either add in Vilkovisky corrections to the $\beta-$functions
or use non-minimal operators in the $\zeta\to\infty$ limit.

The results which we have presented can be extended to combinations of vector
and tensor fields when the operator mixes the two types of field. The general
result for ${\rm Tr}\,E_1$ can still be used for such operators, but the calculations
for ${\rm Tr}\, E_2$ have to be redone to include the cross-terms in the operator.

We have only considered manifolds without boundary, but heat kernel methods
are also important for manifolds with boundary. Some preliminary results
for non-minimal operators on manifolds with boundary can be found in \cite{Avramidi:1997hy}, 
where it was proposed that non-minimal operators may solve the problem of providing
a well-posed boundary value problem for quantum gravity. The methods 
we have used can  be extended to the boundary case and we hope to report on new heat kernel
coefficients in the near future.

\appendix
\section{Tensor identities}

There are relations between products of the tensor $P^{\mu\nu}$ due to
the condition $\hat P^2=\hat P$, and the symmetry
under frame rotations. These relations allow a reduction in the number 
of terms needed to write down explicit forms for the coefficients. 
For example, let $A$ denote the pair of tangent indices $(\mu,\nu)$, 
then ${\rm Tr}\, P_AP_B$ is rotationally invariant, and symmetric
under the interchange of $A$ and $B$. It must take the form
\begin{equation}
{\rm Tr}\, P_AP_B=\beta_1\delta_A\delta_B+\beta_2 t_{AB},\label{tqq}
\end{equation}
where $t_{AB}$ is a totally symmetric tensor in the tangent indices. We can take the symmetric
tensors to have unit trace,
\begin{equation}
t_{\mu_1\mu_2\dots\mu_n}=c_n\,\delta_{(\mu_1\mu_2}\dots\delta_{\mu_{n-1}\mu_n)},
\end{equation}
where
\begin{equation}
c_n=\frac{m!(m/2-1) !}{ 2^m(m/2+n/2-1)!(n/2)!}.
\end{equation}
For example, $c_2=1/m$ and $c_4=3/m(m+2)$. 
Contracting Eq. (\ref{tqq}) with combinations of $\delta^A$ and $\hat p^\mu \hat p^\nu$ gives two relations
involving the spacetime trace $P=P^\mu{}_\mu$,
\begin{eqnarray}
m^2\beta_1+\beta_2&=&{\rm Tr}\,P^2,\\
\beta_1+c_4\beta_2&=&c_2{\rm Tr}\, P.
\end{eqnarray}
Therefore it follows that $\beta_1$ and $\beta_2$ depend only on
${\rm Tr}\, P$ and ${\rm Tr}\, P^2$. A similar argument can be
applied to products of three $P_A$ terms, so that
\begin{lem}\label{lem2}
If $P_A$ is symmetric tensor under tetrad rotations and $\hat P$ is a projection, then
\begin{eqnarray}
&&{\rm Tr}\, P_AP_B=\beta_1\delta_A\delta_B
+\beta_2 t_{AB}.\\
&&{\rm Tr}\, P_AP_BP_C=
\alpha_1\delta_A\delta_B\delta_C
+\alpha_2 \delta_{(A}t_{BC)}+\alpha_3t_{ABC}.
\end{eqnarray}
where $\beta_i$ and $\alpha_i$ only depend on $P=P^\mu{}_\mu$.
\end{lem}
For products involving $P_A$ and the gauge curvature we make use of the
condition $\nabla_\mu P_A=0$, which implies
\begin{equation}
[\nabla_\rho,\nabla_\sigma]P^{\mu\nu}=
[F_{\rho\sigma},P^{\mu\nu}]+2R^{(\mu}{}_{\alpha\rho\sigma}P^{\nu)\alpha}=0.
\label{fcom}
\end{equation}
Terms with the gauge curvature and two $P_A$ factors contributing to $E_1$
can be replaced by Levi-Civita curvature terms, since
\begin{equation}
{\rm Tr}\, P^{\mu\rho}P_\mu{}^\sigma F_{\rho\sigma}=
\frac12{\rm Tr}\, [P^{\mu\rho},P_\mu{}^\sigma] F_{\rho\sigma}
=\frac12{\rm Tr}\, [F_{\rho\sigma},P^{\mu\rho}]P_\mu{}^\sigma
\end{equation}
and Eq. (\ref{fcom}) applies. Together with Lemmas \ref{lem1}
and \ref{lem2}, this allows the $E_1$ coefficient to be expressed in the form \eqref{tre1}.

\section{Normal coordinate expansions}

The $E_2$ coefficients requires the normal coordinate expansions of the vector
operator up to fourth order. These expansions can be found following the procedure
described in Sect. III and are given below.

\begin{eqnarray}
\big(A_{0}^{\mu\nu}\big)^{\lambda}{}_{\tau}&=&-\delta^{\lambda}_{\tau}\delta^{\mu\nu}-\frac{1}{2}\,\zeta\big(\delta^{\mu\lambda}\delta^{\nu}_{\tau}+\delta^{\nu\lambda}\delta^{\mu}_{\tau}\big),\label{3.3.11a}\\
\big(A^{\mu\nu}{}_{\alpha\beta}\big)^{\lambda}{}_{\tau}&=&-\frac{1}{3}\delta^{\lambda}_{\tau}R^{\mu}{}_{\alpha}{}^{\nu}{}_{\beta}-\frac{1}{6}\,\zeta\big( \delta^{\nu}_{\tau}R^{\lambda}{}_{\alpha}{}^{\mu}{}_{\beta} + \delta^{\mu}_{\tau}R^{\lambda}{}_{\alpha}{}^{\nu}{}_{\beta}\big),\label{3.3.11b}\\
\big(A^{\mu\nu}{}_{\alpha\beta\gamma}\big)^{\lambda}{}_{\tau}&=&-\frac{1}{6}\delta^{\lambda}_{\tau}R^{\mu}{}_{\alpha}{}^{\nu}{}_{\beta;\gamma}-\frac{1}{12}\,\zeta\big( \delta^{\nu}_{\tau}R^{\lambda}{}_{\alpha}{}^{\mu}{}_{\beta;\gamma} + \delta^{\mu}_{\tau}R^{\lambda}{}_{\alpha}{}^{\nu}{}_{\beta;\gamma}\big),\label{3.3.11c}\\
\big(A^{\mu\nu}{}_{\alpha\beta\gamma\delta}\big)^{\lambda}{}_{\tau}&=&-\frac{1}{20}\,\delta^{\lambda}_{\tau}\,R^{\mu}{}_{\alpha}{}^{\nu}{}_{\beta;\gamma\delta} -\frac{1}{15}\,\delta^{\lambda}_{\tau}\, R^{\mu}{}_{\alpha\rho\beta}\,R^{\nu}{}_{\gamma}{}^{\rho}{}_{\delta}\nonumber\\
&& -\frac{1}{10}\,\zeta\big( \frac{1}{4}\,\delta^{\nu}_{\tau}\,R^{\lambda}{}_{\alpha}{}^{\mu}{}_{\beta;\gamma\delta} + \frac{1}{4}\,\delta^{\mu}_{\tau}\,R^{\lambda}{}_{\alpha}{}^{\nu}{}_{\beta;\gamma\delta}\nonumber\\
&&\qquad + \frac{1}{3}\,\delta^{\nu}_{\tau}\,R^{\lambda}{}_{\alpha\sigma\beta}\,R^{\mu}{}_{\gamma}{}^{\sigma}{}_{\delta}
+ \frac{1}{3}\,\delta^{\mu}_{\tau}\,R^{\lambda}{}_{\alpha\sigma\beta}\,R^{\nu}{}_{\gamma}{}^{\sigma}{}_{\delta}
\big),\label{3.3.11d}\\
\big(B^{\mu}{}_{\alpha}\big)^{\lambda}{}_{\tau}&=&\frac{2}{3}\delta^{\lambda}_{\tau}R^{\mu}{}_{\alpha} -\frac{2}{3}R_{\alpha\tau}{}^{\lambda\mu}-\frac{2}{3}R_{\alpha}{}^{\mu\lambda}{}_{\tau} +\frac{1}{3}\,\zeta\,\delta^{\mu\lambda}\,R_{\tau\alpha},\label{3.3.11e}\\
\big(B^{\mu}{}_{\alpha\beta}\big)^{\lambda}{}_{\tau}&=&\delta^{\lambda}_{\tau}\Big(\frac{1}{3}R^{\mu}{}_{\alpha;\beta} +\frac{1}{6}R^{\mu}{}_{\alpha\beta}{}^{\sigma}{}_{;\sigma} +\frac{1}{12}R_{\alpha\beta}{}^{;\mu}\Big) -\frac{1}{3}R_{\alpha\tau}{}^{\lambda\mu}{}_{;\beta}\nonumber\\
&&\hspace{-12pt}-\frac{1}{3}R_{\alpha}{}^{\mu\lambda}{}_{\tau;\beta} +\frac{1}{6}R_{\alpha}{}^{\lambda}{}_{\beta\tau}{}^{;\mu} +\frac{1}{6}R_{\alpha}{}^{\lambda}{}_{\beta}{}^{\mu}{}_{;\tau} -\frac{1}{6}R_{\alpha}{}^{\mu}{}_{\beta\tau}{}^{;\lambda}\nonumber\\
&&+\frac{1}{12}\,\zeta\,\delta^{\mu\lambda}\big(2R_{\tau\alpha;\beta}+R_{\alpha\beta;\tau}\big),\label{3.3.11f}\\
\big(B^{\mu}{}_{\alpha\beta\gamma}\big)^{\lambda}{}_{\tau}&=&\delta^{\lambda}_{\tau}\,\Big(\frac{1}{10}R^{\mu}{}_{\alpha;\beta\gamma}
+\frac{1}{20}R^{\mu}{}_{\alpha\beta}{}^{\sigma}{}_{;\sigma\gamma}
+\frac{1}{20}R^{\mu}{}_{\alpha\beta}{}^{\sigma}{}_{;\gamma\sigma}
+\frac{1}{40}R_{\alpha\beta}{}^{;\mu}{}_{\gamma}
\nonumber\\
&&+\frac{1}{40}R_{\alpha\beta;\gamma}{}^{\mu}
-\frac{8}{45}R^{\mu}{}_{\alpha\beta}{}^{\sigma}R_{\sigma\gamma}
-\frac{4}{45}R^{\mu}{}_{\lambda\sigma\alpha}R^{\lambda}{}_{\beta}{}^{\sigma}{}_{\gamma} \Big) \nonumber\\
&&+\frac{1}{10}R^{\lambda\mu}{}_{\tau\gamma;\alpha\beta} +\frac{1}{10}R^{\lambda}{}_{\tau\mu\gamma;\alpha\beta}+\frac{1}{20}R^{\lambda}{}_{\beta\tau\gamma}{}^{;\mu}{}_{\alpha} +\frac{1}{20}R^{\lambda}{}_{\beta}{}^{\mu}{}_{\gamma;\tau}{}^{\alpha} \nonumber\\
&&+\frac{1}{20}R^{\lambda}{}_{\beta\tau\gamma;\alpha}{}^{\mu} +\frac{1}{20}R^{\lambda}{}_{\beta}{}^{\mu}{}_{\gamma;\alpha\tau}-\frac{1}{20}R^{\mu}{}_{\beta\tau\gamma}{}^{;\lambda}{}_{\alpha} -\frac{1}{20}R^{\mu}{}_{\beta\tau\gamma;\alpha}{}^{\lambda} \nonumber\\
&&+\frac{2}{45}R^{\lambda}{}_{\alpha}{}^{\mu\nu}R_{\tau\beta\nu\gamma} +\frac{2}{45}R^{\lambda}{}_{\alpha\tau}{}^{\nu}R^{\mu}{}_{\beta\nu\gamma}-\frac{8}{45}R^{\lambda}{}_{\alpha}{}^{\nu}{}_{\beta}R^{\mu}{}_{\nu\tau\gamma} \nonumber\\
&&-\frac{8}{45}R^{\lambda}{}_{\alpha}{}^{\nu}{}_{\beta}R^{\mu}{}_{\gamma\tau\nu} -\frac{2}{45}R^{\lambda\nu\mu}{}_{\beta}R_{\tau\alpha\nu\gamma}+\frac{8}{45}R^{\lambda\nu}{}_{\tau\beta}R^{\mu}{}_{\alpha\nu\gamma}\nonumber\\
&&-\frac{4}{45}R^{\lambda\mu\nu}{}_{\beta}R_{\tau\alpha\nu\gamma} +\frac{2}{15}R^{\lambda}{}_{\tau}{}^{\nu}{}_{\beta}R^{\mu}{}_{\alpha\nu\gamma}+\frac{1}{9}\,\zeta\,R^{\mu}{}_{\alpha}{}^{\lambda}{}_{\beta}\,R_{\tau\gamma}\nonumber\\
&&-\zeta\,\delta^{\mu\lambda}\Big(-\frac{1}{20}\,R_{\tau\gamma;\alpha\beta}-\frac{1}{40}\,R_{\beta\gamma;\tau\alpha}-\frac{1}{4}\,R_{\beta\gamma;\alpha\tau}\nonumber\\
&&-\frac{8}{45}\,R^{\rho}{}_{\alpha\beta\sigma}\,R_{\tau\gamma\rho}{}^{\sigma}-\frac{1}{45}\,R^{\rho}{}_{\alpha\beta\sigma}\,R_{\rho\tau\gamma}{}^{\sigma} \Big) ,\label{3.3.11g}
\end{eqnarray}
\begin{eqnarray}
\big( C_0\big)^{\lambda}{}_{\tau}&=&Q^{\lambda}{}_{\tau}+\frac{1}{3}R^{\lambda}{}_{\tau} +\frac{1}{3}\,\zeta\,R^{\lambda}{}_{\tau},\label{3.3.12a}\\
\big( C_{\alpha}\big)^{\lambda}{}_{\tau}&=&Q^{\lambda}{}_{\tau;\alpha}+\frac{1}{6}R^{\lambda}{}_{\tau;\alpha} 
+\frac{1}{4}R^{\lambda\beta}{}_{\tau\alpha;\beta} -\frac{1}{12}R^{\lambda}{}_{\alpha;\tau}\nonumber\\
&&-\frac{1}{6}R^{\lambda}{}_{\tau\alpha}{}^{\beta}{}_{;\beta} +\frac{1}{12}R^{\lambda}{}_{\alpha\tau}{}^{\beta}{}_{;\beta} +\frac{1}{12}R_{\tau\alpha}{}^{;\lambda}\nonumber\\
&&+\frac{1}{6}\,\zeta\,\big(R_{\alpha}{}^{\lambda}{}_{;\tau}+R_{\alpha\tau}{}^{;\lambda}+R_{\tau}{}^{\lambda}{}_{;\alpha}\big),\label{3.3.12b}\\
\big( C_{\alpha\beta}\big)^{\lambda}{}_{\tau}&=&\frac{1}{2}\,Q^{\lambda}{}_{\tau;\alpha\beta}+\frac{1}{6}R^{\lambda}{}_{\alpha\sigma\beta}Q^{\sigma}{}_{\tau} -\frac{1}{6}R^{\sigma}{}_{\alpha\tau\beta} Q^{\lambda}{}_{\sigma} +\frac{1}{20}R^{\lambda}{}_{\tau;\alpha\beta}\nonumber\\
&&-\frac{1}{40}R^{\lambda}{}_{\alpha;\tau\beta}-\frac{1}{40}R^{\lambda}{}_{\alpha;\beta\tau}+\frac{1}{40}R_{\tau\alpha}{}^{;\lambda}{}_{\beta} +\frac{1}{40}R_{\tau\alpha;\beta}{}^{\lambda}\nonumber\\
&&+\frac{3}{40}R^{\lambda\mu}{}_{\tau\alpha;\mu\beta}+\frac{3}{40}R^{\lambda\mu}{}_{\tau\alpha;\beta\mu} +\frac{1}{20}R^{\lambda}{}_{\tau}{}^{\mu}{}_{\alpha;\mu\beta} +\frac{1}{20}R^{\lambda}{}_{\tau}{}^{\mu}{}_{\alpha;\beta\mu}\nonumber\\
&&+\frac{1}{40}R^{\lambda}{}_{\alpha\tau}{}^{\mu}{}_{;\mu\beta} +\frac{1}{40}R^{\lambda}{}_{\alpha\tau}{}^{\mu}{}_{;\beta\mu} +\frac{1}{20}\nabla^2 R^{\lambda}{}_{\alpha\tau\beta} +\frac{1}{40}R^{\lambda}{}_{\alpha}{}^{\mu}{}_{\beta;\tau\mu}  \nonumber\\
&&+\frac{1}{40}R^{\lambda}{}_{\alpha}{}^{\mu}{}_{\beta;\mu\tau}-\frac{1}{40}R_{\tau\alpha\mu\beta}{}^{;\mu\lambda} -\frac{1}{40}R_{\tau\alpha\mu\beta}{}^{;\lambda\mu} - \frac{1}{45}R^{\lambda}{}_{\alpha\tau\mu}R^{\mu}{}_{\beta}\nonumber\\
&& +\frac{4}{45}R^{\lambda}{}_{\alpha\mu\beta}R^{\mu}{}_{\tau} -\frac{1}{15}R^{\lambda\mu}R_{\tau\alpha\mu\beta} -\frac{1}{5}R^{\lambda}{}_{\mu\tau\alpha}R^{\mu}{}_{\beta}-\frac{8}{45}R^{\lambda}{}_{\tau\mu\alpha}R^{\mu}{}_{\beta}\nonumber\\
&& -\frac{1}{15}R^{\lambda}{}_{\alpha\mu\nu}R_{\tau}{}^{\mu\nu}{}_{\beta} + \frac{1}{15}R^{\lambda}{}_{\alpha\mu\nu}R_{\tau\beta}{}^{\mu\nu} +\frac{1}{9}R^{\lambda}{}_{\mu\tau\nu}R^{\mu}{}_{\alpha}{}^{\nu}{}_{\beta}\nonumber\\
&& - \frac{1}{15}R^{\lambda}{}_{\mu\nu\alpha}R_{\tau\beta}{}^{\mu\nu} +\frac{1}{15}R^{\lambda}{}_{\mu\nu\alpha}R_{\tau}{}^{\mu\nu}{}_{\beta}\nonumber\\
&&-\zeta\Big(-\frac{1}{20}\,R_{\tau\alpha}{}^{;\lambda}{}_{\beta}-\frac{1}{40}\,R_{\alpha\beta;\tau}{}^{\lambda} -\frac{1}{40}\,R_{\alpha\beta}{}^{;\lambda}{}_{\tau}-\frac{1}{20}\,R_{\tau\beta;\alpha}{}^{\lambda}\nonumber\\
&&-\frac{1}{40}\,R^{\lambda}{}_{\beta;\tau\alpha}-\frac{1}{40}\,R^{\lambda}{}_{\beta;\alpha\tau}-\frac{1}{20}\,R_{\tau}{}^{\lambda}{}_{;\alpha\beta}-\frac{1}{40}\,R_{\beta}{}^{\lambda}{}_{;\tau\alpha}-\frac{1}{40}\,R_{\beta}{}^{\lambda}{}_{;\alpha\tau}\nonumber\\
&&-\frac{1}{45}\,R^{\rho\lambda}{}_{\beta}{}^{\sigma}\,R_{\rho\tau\alpha\sigma} +\frac{8}{45}\,R^{\rho\lambda}{}_{\beta}{}^{\sigma}\,R_{\rho\sigma\alpha\tau}-\frac{1}{45}\,R^{\rho}{}_{\alpha}{}^{\lambda\sigma}\,R_{\rho\tau\beta\sigma}\nonumber\\
&&+\frac{8}{45}\,R^{\rho}{}_{\alpha}{}^{\lambda\sigma}\,R_{\rho\sigma\beta\tau}-\frac{1}{45}\,R^{\rho}{}_{\alpha\beta}{}^{\sigma}\,R_{\rho\tau}{}^{\lambda}{}_{\sigma}+\frac{8}{45}\,R^{\rho}{}_{\alpha\beta}{}^{\sigma}\,R_{\rho\sigma}{}^{\lambda}{}_{\tau}\nonumber\\
&&-\frac{1}{9}\,R^{\lambda}{}_{\alpha}{}^{\rho}{}_{\beta}\,R_{\tau\rho}\Big).\label{3.3.12c}
\end{eqnarray}
Strictly speaking the results in \eqref{3.3.11b}, \eqref{3.3.11f} and \eqref{3.3.12c} should be symmetrized in $\alpha$ and $\beta$, the results in \eqref{3.3.11c} and \eqref{3.3.11g} should be symmetrized in $\alpha,\beta,\gamma$, and the result in \eqref{3.3.11d} should be symmetrized in $\alpha,\beta,\gamma,\delta$; however as these expressions are contracted with symmetric terms in our general results we can ignore this symmetrization for brevity. 

\section{Untraced $E_1$ coefficients}\label{appuntraced}

Here we give the expression for the untraced $E_1$ coefficient for the Yang-Mills non-minimal operator. We do not give the analogous result for gravity as the expression is somewhat more lengthy.
\begin{eqnarray}
\big(E_1\big)^{a\mu}{}_{b\nu}&=&\alpha\,Q^{a\mu}{}_{b\nu}+\beta\,Q^{a}{}_{\nu b}{}^{\mu}+\beta\,Q^{a\lambda}{}_{b\lambda}\,\delta^{\mu}_{\nu} +\gamma\,\delta^{a}_{b}\,R^{\mu}{}_{\nu}+\delta\,\delta^{a}_{b}\,R\,\delta^{\mu}_{\nu}\nonumber\\
&&\qquad +\varepsilon\,g\,f^{a}{}_{ b c} F^{c\,\mu}{}_{\nu}.\label{C1}
\end{eqnarray}
The coefficients in this expression are given by
\begin{eqnarray}
\alpha&=&\frac{(3m+6)\zeta+4m+4}{m(m^2-4)\zeta}\,(1+\zeta)^{-m/2}\nonumber\\
&&\qquad-\,\frac{(6-3m-2m^2+m^3)\zeta+4m+4}{m(m^2-4)\zeta},\label{C2}\\
\beta&=&-\frac{(m\zeta+2\zeta+4)}{m(m^2-4)\zeta}\;(1+\zeta)^{-m/2}-\frac{(m\zeta-2\zeta-4)}{m(m^2-4)\zeta},\label{C3}\\
\gamma&=&-\frac{m(m+2)\zeta^2+(m+2)(m+6)\zeta+12m}{3m(m^2-4)\zeta}\;(1+\zeta)^{-m/2}\nonumber\\
&&\qquad-\,\frac{(5m^2-8m-12)\zeta-12m}{3m(m^2-4)\zeta},\label{C4}\\
\delta&=&\frac{m(m+2)\zeta^2+(m+2)(m+6)\zeta+24}{6m(m^2-4)\zeta}\;(1+\zeta)^{-m/2}\nonumber\\
&&\qquad+\frac{(m^3-m^2-12)\zeta-24}{6m(m^2-4)\zeta},\label{C5}\\
\varepsilon&=&-\frac{(3m-8)\zeta-8}{m(m-2)\zeta}-\;\frac{m\zeta+8}{m(m-2)\zeta}\;(1+\zeta)^{1-m/2}.\label{C6}
\end{eqnarray}
These results agree with those found by Gusynin and Kornyak~\cite{gusynin1999complete} except that their expression for $W_{ij}$ should only include the Yang-Mills gauge connection if we are to have agreement. We give the minimal operator and $m=4$ limits in table~\ref{tableappC}. The minimal operator has the standard form of $\displaystyle{\frac{1}{6}RI-Q}$. 

\begin{table}[htb]
\caption{\label{tableappC}Limiting form of coefficients in the untraced heat kernel coefficient $\big(E_1\big)^{a\mu}{}_{b\nu}$ for 
Yang-Mills theory with minimal operators $\zeta=0$ or spacetime dimension $m=4$. }
\centering
\begingroup
\renewcommand{\arraystretch}{1.3}
\begin{tabular}{>{$}l<{$}>{$\displaystyle}c<{$}>{$\displaystyle}c <{$}}
\hline\noalign{\smallskip}
{\rm Term}&\zeta=0&m=4\\
\noalign{\smallskip}\hline\noalign{\smallskip}
\alpha&-1&-\,\frac{\zeta^2+6\zeta+4}{4(1+\zeta)^2}\\
\beta&0 &-\,\frac{\zeta^2}{24(1+\zeta)^2} ,\\
\gamma&0&-\frac{\zeta(3\zeta+4)}{12 (1+\zeta)^2} ,\\
\delta&\frac{1}{6} &\frac{3 \zeta^2+6 \zeta+4}{24 (1+\zeta)^2} ,\\
\varepsilon&0&-\,\frac{\zeta}{2 (1+\zeta)} .\\
\noalign{\smallskip}\hline
\end{tabular}
\endgroup
\end{table}

\begin{acknowledgments}
DJT is grateful to S. A. Fulling for providing him with the references \cite{fulling1992kernel} and \cite{fulling1992kernel2}.
IGM is supported by the Science and Technology Facilities Council Consolidated Grant ST/J000426/1.
\end{acknowledgments}


\end{document}